\input{psfig.sty}

\documentclass[galaxies,review,accept,oneauthor,pdftex]{Definitions/mdpi} 

\usepackage{soul}
\usepackage{upgreek}
\usepackage{enumitem}
\usepackage{subfigure}
\usepackage{booktabs}
\usepackage{multirow}
\usepackage{longtable}
\usepackage{microtype}
\newcommand\myurl[1]{\changeurlcolor{black}\url{#1}\changeurlcolor{blue}}

\usepackage{subfigure}
\makeatletter
\renewcommand{\@thesubfigure}{\normalsize(\textbf{\alph{subfigure}})}
\makeatother

\firstpage{1}
\pubvolume{9}
\issuenum{3}
\articlenumber{60}
\pubyear{2021}
\copyrightyear{2021}
\externaleditor{Academic Editor:
Fulai Guo} 
\datereceived{8 July 2021}
\dateaccepted{26 August 2021}
\datepublished{31 August 2021}
\hreflink{https://doi.org/10.3390/\linebreak galaxies9030060} 
\pdfoutput=1

\usepackage[utf8]{inputenc}
\usepackage{amsthm}

\theoremstyle{mdpi}
\newcounter{thm}
\newcounter{ex}
\newcounter{re}
\newtheorem{Definitions/Lemma}[thm]{Lemma}
\newtheorem{Definitions/Corollary}[thm]{Corollary}
\newtheorem{Definitions/Proposition}[thm]{Proposition}

\theoremstyle{mdpidefinition}
\newtheorem{Definitions/Characterization}[thm]{Characterization}
\newtheorem{Definitions/Property}[thm]{Property}
\newtheorem{Definitions/Problem}[thm]{Problem}
\newtheorem{Definitions/Example}[ex]{Example}
\newtheorem{Definitions/ExamplesandDefinitions}[ex]{Examples and Definitions}
\newtheorem{Definitions/Remark}[re]{Remark}
\newtheorem{Definitions/Definition}[thm]{Definition}
\def\aap{A\&A}

\def\aap{Astron. Astrophys.}

\def\apj{ApJ}

\def\apjs{ApJS}

\def\mnras{MNRAS}

\def\apj{Astrophys. J.}

\def\apjs{Astrophys. J. Supp.}

\def\mnras{Mon. Not. R. Astron. Soc.}

\usepackage{amsmath}

\def\nvphantom{\v@true\h@false\nph@nt}
\def\nhphantom{\v@false\h@true\nph@nt}
\def\nphantom{\v@true\h@true\nph@nt}
\def\nph@nt{\ifmmode\def\next{\mathpalette\nmathph@nt}%
\else\let\next\nmakeph@nt\fi\next}
\def\nmakeph@nt#1{\setbox\z@\hbox{#1}\nfinph@nt}
\def\nmathph@nt#1#2{\setbox\z@\hbox{$\m@th#1{#2}$}\nfinph@nt}
\def\nfinph@nt{\setbox\tw@\null
\ifv@ \ht\tw@\ht\z@ \dp\tw@\dp\z@\fi
\ifh@ \wd\tw@-\wd\z@\fi \box\tw@}

\Title{On the Origin and Evolution of the Intra-Cluster Light: A Brief Review of the Most Recent Developments}

\TitleCitation{On the Origin and Evolution of the Intra-Cluster Light: A Brief Review of the Most Recent Developments}

\Author{Emanuele Contini
}

\AuthorNames{Emanuele Contini}
\AuthorCitation{Contini, E.}

\address[1]{%
School of Astronomy and Space Science, Nanjing University, Nanjing 210093, China; {emanuele.contini82@gmail.com}}

\abstract{Not all the light in galaxy groups and clusters comes from stars that are bound to galaxies.
A significant fraction of it constitutes the so-called intracluster or diffuse light (ICL), a low surface brightness
component of groups/clusters generally found in the surroundings of the brightest cluster galaxies and
intermediate/massive satellites. In this review, I will describe the mechanisms responsible for its formation
and evolution, considering the large contribution given to the topic in the last decades by both the theoretical
and observational sides. Starting from the methods that are commonly used to isolate the ICL, I will address
the remarkable problem given by its own definition, which still makes the comparisons among different studies
not trivial, to conclude by giving an overview of the most recent works that take advantage of the ICL as a
luminous tracer of the dark matter distribution in galaxy groups and clusters.}


\begin{document}
\section{Introduction}\label{sec:intro}

Since its discovery by \citet{zwicky37}, who suggested for the first time the existence of a component made of stars that ``float'' between cluster galaxies, the Intra-Cluster Light (hereafter ICL) has been gradually considered an important luminous component in galaxy groups and clusters. Zwicky's suggestion was an extraordinary prediction, since almost a century later we consider the ICL to be a diffuse component in clusters that is made of stars
not gravitationally bound to any galaxy member of the cluster ({the history
of the ICL, in particular that of the last century, goes beyond the scope of this review. For a historical survey, I refer the reader to, e.g., the review by \citet{montes19rev}}). Most of the ICL in clusters is considered to be somehow associated with the brightest cluster galaxy (hereafter BCG), which is the galaxy that resides in the center of clusters (for the sake of simplicity, I will call BCGs also those residing in the center of galaxy groups), while a  smaller fraction of the total is found, and also predicted, to be around intermediate and massive satellites (\mbox{e.g., \cite{gonzalez13,contini14,presotto14,contini18}}). The BCG and its associated ICL are commonly taken as a whole system (e.g., \cite{arnaboldi12,edwards16,kravtsov18,zhang19,demaio20,montes21}, just to quote a few of the most recent observational studies), given the actual observational difficulty to separate them (I will fully address this point in Section \ref{sec:definitions}).

The definition of the ICL is probably the most serious issue for observational\linebreak \mbox{studies (\cite{rudick11,cui14,jimenez16,montes18,tang18,zhang19})}, while it is much less complex in theoretical ones. A non-unique definition of the ICL makes comparisons among studies very complicated, and this applies not only between theoretical and observational works but also between the same ``class''. In  hydrosimulations (e.g., \cite{dolag10,rudick11}), it is quite easy to define the ``actual'' ICL simply because of the information of each particle that the user can have access to, and in semi-analytic models (e.g., \cite{somerville08,henriques10,guo11,contini14}), it is straightforward given that fact that the amount of ICL is the result of a few numerical equations. Observers cannot rely on the information available with numerical models but observe just the light coming from an area in the sky. There are mainly two observational methods, which I will fully address in the next section, that are used to define the ICL. A common method frequently used, especially in the past, is to separate BCG and ICL according to a given cut in surface brightness and after removing all the other contributions (e.g, \cite{feldmeier04,zibetti05,furnell21}). The ICL is then defined as all the light fainter than the cut. A second method, more common in the recent past, is to use functional forms to distribute the BCG + ICL light such as a double/triple Sersic \cite{sersic68}, exponential or composite profiles (e.g., \cite{gonzalez05,seigar07,donzelli11,giallongo14,cooper15,alamo17,durret19,kluge21}). The assumption is that the ICL contribution starts at the distance when the light profile drops and follows a plateau. Clearly, each of them has pros and cons, which will be discussed in Section \ref{sec:definitions} by giving a few examples. It must be stressed, however, that although theoretical models can be more precise in both detecting the ICL and in separating it from the BCG, observational techniques should be those to follow in order to have a fair comparison of the results coming from theory and observations. Moreover, as most of the studies have done and still do, the easiest approach would be to not separate the two components at all and study their properties as though they were a \mbox{single system.}

Another possible way to detect the ICL, for near objects, is given by the globular cluster (GC) population. GC are considered good tracers of the ICL since {the contamination
of the sample from background stars and foreground galaxies is almost zero}. They have been extensively used in nearby clusters such as {Fornax \cite{schuberth08,dabrusco16,cantiello20}, Virgo \cite{lee10,durrell14,ko17,longobardi18}, \mbox{Perseus \cite{harris20}}, Coma \cite{peng11,madrid18}, A1185 \cite{west11}, Abell 1689 \cite{alamo17}, and others}. As we will see below, stellar stripping is an important mechanism to form the ICL, and it is also supposed to strip GCs from their parent galaxies, forming the so-called intracluster CGs. In close objects, the ICL can be detected also using Planetary Nebulae (PNe). A recent example is given by \citet{longobardi15} (but see also references therein), who studied a sample of 287 PNe around the BCG M 87 in Virgo A and have been able to distinguish the stellar halo from the ICL and study the two components separately. This result could be achieved because the velocity distribution of PNe is bimodal, with a narrow component centered on the BCG and a broader one (off-centered) that can be associated with halo and ICL. Attempts to see the transition between the BCG and the ICL have been made even earlier and by using dynamical measurements from spectroscopy. For instance, \citet{kelson02} found that the velocity dispersion of stars in the cluster A2199 decreases from the initial central value (BCG dominated), to rise again after a few kpc, almost reaching the velocity dispersion of the cluster at larger radii. The authors suggested that the stars in the halo of the cD galaxy trace the potential of the cluster, and that the kinematics of the ICL stars can be used to constrain the mass profile of the cluster (see also \citet{bender15}).

In the last couple of decades, a few physical mechanisms responsible for the formation of the ICL have been proposed, and even a combination of them in theoretical models. The most important ones can be listed as follows: (a) violent relaxation processes during galaxy mergers \cite{monaco06,murante07,gerhard07,contini14,burke15,groenewald17,jimenez18,jimenez19}; (b) disruption of dwarf galaxies \cite{purcell07,murante07,conroy07,martel12,giallongo14,annunziatella16,morishita17,raj20}; (c) tidal stripping of intermediate and massive galaxies \cite{rudick09,rudick11,contini14,demaio15,demaio18,montes18,jimenez18,jimenez19}; (d) pre-processing and accretion from other objects \cite{willman04,mihos05,rudick06,sommer-larsen06,contini14}; (e) {in situ} star formation \cite{puchwein10}. In recent years, among the above-mentioned physical processes, only mergers, stellar stripping and pre-processing have been qualified as main channels. Indeed, {in situ} star formation, which was originally proposed by \citet{puchwein10}, has been ruled out by the observed evidence (e.g., \cite{melnick12}) that only 1\% of the ICL can be attributed to this channel. Similarly, disruption of dwarf galaxies has been shown to be a secondary channel by several works, both theoretical and observational (e.g., \cite{montes14,contini14,demaio15,montes18}). This particular channel appears to be important in terms of the number of galaxies involved in the process, rather than the effective mass that they provide to the ICL. In a way, mergers and pre-processing have been classified as less important than stripping of intermediate/massive galaxies by several authors (references above). For instance, \citet{contini14} showed that mergers can account for no more than 15--20\% of the total ICL and that pre-processing is particularly important only in massive clusters, for which it can contribute up to 30\% of the total ICL mass. In Sections \ref{sec:formation} and \ref{sec:properties}, I will provide a more detailed picture, supported by observed evidence, that stellar stripping is the most important formation channel for the ICL.

Indeed, depending on the particular channel, properties of the ICL (predicted by models or observed), such as metallicity and color distributions, can appear rather different. By looking at these two properties, it is possible to discern which, among the above quoted mechanisms, is the main responsible mechanism for the formation of the ICL. In the last few years, many authors focused on the age, color and metallicity gradients of BCG + ICL systems \cite{zibetti05,krick06,krick07,rudick10,toledo11,montes14,demaio15,morishita17,iodice17,harris17,demaio18,montes18,contini19,zhang19,spavone20,edwards20,gu20} and in particular on their shape. It is reasonable to argue that, if mergers are the main responsible mechanisms, we would expect a flat gradient, i.e., no gradient, simply because mergers would tend to mix up the two components (see discussion in \cite{contini18,contini19}). On the contrary, a more gradual formation of the ICL via stellar stripping during the evolution in time of the clusters would end up in some form of color/metallicity gradient, positive or negative depending on the stars that are contributing to the \mbox{ICL \cite{demaio15,morishita17,iodice17,demaio18,montes18,contini19}.} Currently, most of the observations (and theoretical predictions) are favoring a picture where stellar stripping dominates the formation of the ICL because in most cases those properties do show a gradient (see references above).

Among the most important properties of the ICL, its radial distribution is surely one of them and it can shed some light on two delicate topics: The dynamical state of clusters (\mbox{e.g., \cite{gerhard07,doherty09,ventimiglia11}}) and the dark matter (hereafter DM) distribution. Recently, there have been many attempts (e.g., \cite{pillepich18,montes19,zhang19,alonso20,kluge20,contini20,contini21,sampaio21,poliakov21,deason21,gonzalez21}) to link the ICL with the DM in clusters. For instance, \citet{montes19}, with a sample of six clusters from the Hubble Frontier Fileds \cite{lotz17}, investigated the ICL and DM distributions by means of the modified Hausdorff distance, a method that connects the two distributions and quantifies how far the two components are from each other. These authors found that the average distance between ICL and DM is around 25 kpc within the first 140 kpc from the center, making the ICL a more reliable tracer of the mass distribution than, e.g., X-ray  emissions ({X-ray emission,
however, has been important as well for the detection of the ICL. Indeed, some studies on Virgo (e.g., \cite{hou17}) and Fornax (e.g., \cite{jin19}) have shown that the X-ray band can be a sensitive and independent way to trace the ICL, although, given the current amount of data available, these studies would be limited to near groups/clusters.}). A similar conclusion has been reached by other authors (e.g., \cite{kluge20,alonso20,contini20,contini21}) with different methods. The ICL is then becoming an observable tracer of the DM distribution and this approach, i.e., linking ICL and DM, is surely a promising way to learn more about the dynamical evolution of galaxy clusters.

This review is structured as follows: in Section \ref{sec:definitions}, I will give a brief summary of the main important methods to define the ICL, with a few examples. Section \ref{sec:formation} will address in detail all the mechanisms that are responsible for the formation of the ICL, and \mbox{Section \ref{sec:properties}} its most important properties, while Section \ref{sec:tracer} will address the most promising side in the study of the ICL, i.e., as a tracer of the DM. Finally, in Section \ref{sec:conclusions}, I will briefly summarize the main topics addressed and give some concluding remarks.

\section{Definition of the ICL}\label{sec:definitions}

\begin{figure}[H]
\includegraphics[width=12 cm]{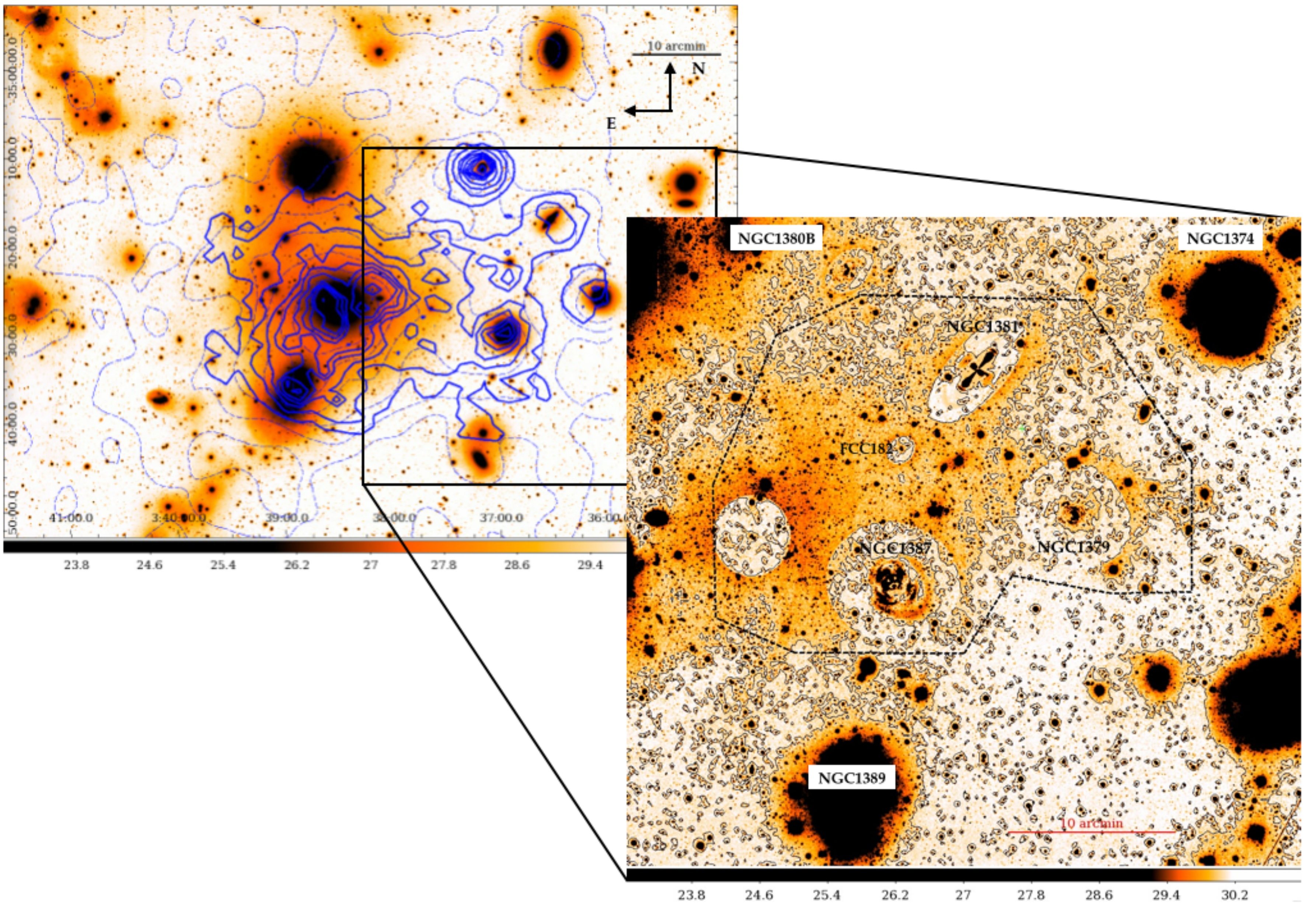}\\
\includegraphics[width=12 cm]{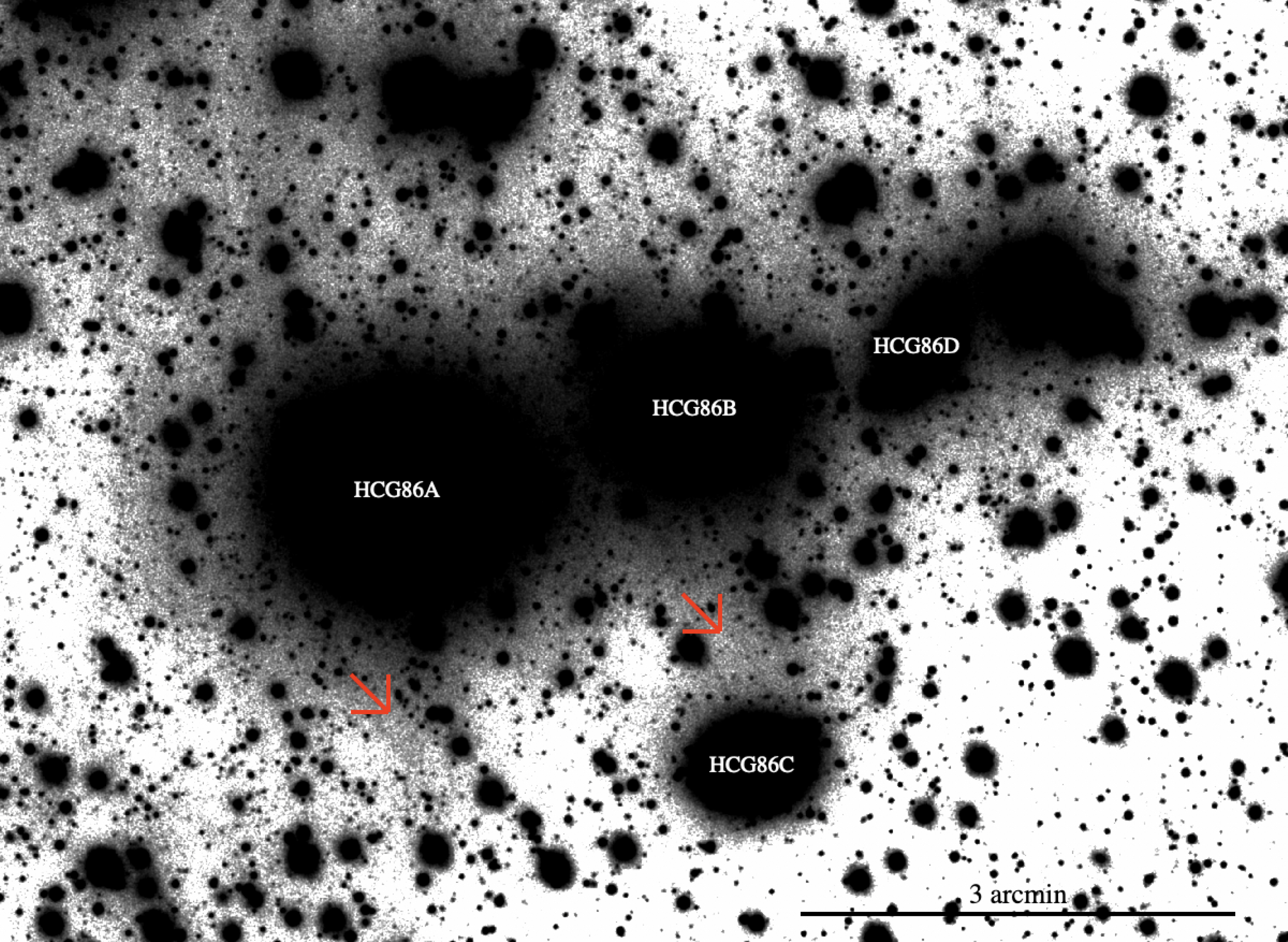}
\caption{Top panel: An example of ICL in the Fornax cluster in {\em r}-band. Left panel: The blue contours are the spatial distribution of the globular clusters derived by \cite{dabrusco16,cantiello18}. Right panel: zoom-in of the west side where it is possible to see how the ICL distributes in the intra-cluster space. Credit: \mbox{\citet{iodice17}}. Bottom panel: a recent example of ICL (in this case intragroup light) in the HCG \mbox{86 group}. The grey detection indicated also by red arrows is the detection of the diffuse light. Credit: \mbox{\citet{ragusa21}}.}
\label{fig:ragusa}
\end{figure}

{The ICL
is an important component in galaxy clusters that fills the space between galaxies, but it is not easy to fully detect. The literature has plenty of beautiful images that can help to understand the importance of such a component. Two of them are shown in Figure \ref{fig:ragusa}. The top panel of Figure \ref{fig:ragusa} (from \citet{iodice17}) displays the ICL ({\em r}-band) on the west side of the Fornax cluster. In the top-right panel of the figure, the authors show a map of the ICL,
which is a residual image after having masked the contribution coming from the early-type galaxies present in the area. The bottom panel (from \citet{ragusa21}) displays a deep VST image in the {\em g} band of the HCG 86 group (but see also \cite{gu20} and references therein for the Coma cluster, and \cite{mihos17} and references therein for the Virgo cluster). }

As briefly discussed in Section \ref{sec:intro}, defining the ICL, i.e., separating it from the BCG to which it is associated, is not an easy task. There are, however, a variety of observational and numerical methods very different from each other, which often result in different amounts of ICL and thus result in indirect comparisons between studies. On the observational side, the most common method used is the ``isophotal limit cut-off'' (e.g., \cite{zibetti05}) in surface brightness, which assumes that the ICL is the remaining component below the cut, after having removed the contribution coming from other sources, such as sky and satellite galaxies. Another method that is increasingly being used lastly relies, instead, on profile fittings with functional forms to model the BCG + ICL component (e.g., \cite{gonzalez05}).

On the numerical side, given the large amount of information available (not achievable in observations), the ICL can be defined, e.g., by taking advantage of the dynamical information provided by the simulation (e.g., \cite{dolag10}), or by using some binding energy definitions in order to separate all the stars that are not bound to any galaxy (e.g., \cite{murante07}). Of course, numerical techniques can mimic the observational methods, i.e., surface brightness cut and profile fittings can be reproduced in simulations (see \cite{rudick11,cui14,tang18} and many others). Below, I will summarize, by providing some examples, the most common observational and theoretical approaches in separating the BCG from its associated ICL.

\subsection{Observational Methods}\label{sec:obs_meth}

As mentioned above, the two most common methods to define the ICL observationally rely only on the light that one can observe. The easiest, but not trivial, way to separate the ICL from the rest (after having removed the contribution from other sources, BCG included) is by assuming a cut (different depending on the band) in the surface brightness. Clearly, the cut that can be chosen is quite arbitrary, and there is no value decided a priori that can be adopted as the standard one. This method has been used, and still is, by several authors (see references above). For example, \citet{zibetti05} analyzed the spatial distribution and color of the ICL by stacking almost 700 galaxy clusters in the redshift range $0.2<z<0.3$ taken from the first release of the Sloan Digital Sky Survey. The authors traced the surface brightness profile of the ICL out to 700 kpc and found that it ranges from 27.5 mag/arcsec$^2$ at 100 kpc to down to around 32 mag/arcsec$^2$ at 700 kpc in the $r-$band. The contribution of the ICL was found to increase with the distance from the center and later studies (\mbox{e.g., \cite{gonzalez07,iodice16,iodice17,montes18}} and others) confirmed it.

\begin{figure}[H]
\includegraphics[width=13 cm]{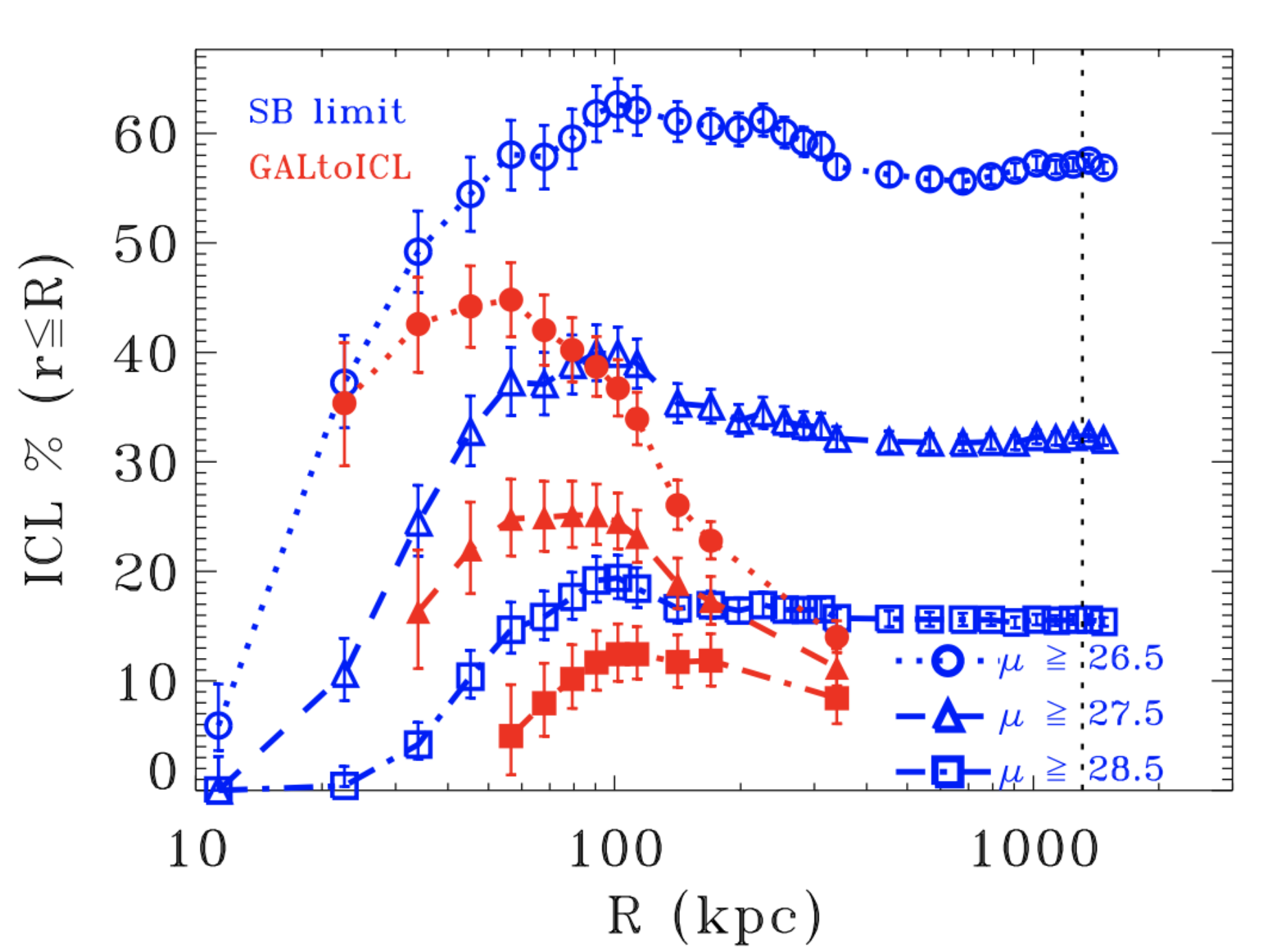}
\caption{The ICL fraction as a function of distance from the center for different surface brightness cuts and different ICL measurements methods. Blue symbols and lines refer to the standard surface brightness method, while red symbols and lines refer to the method developed in \citet{presotto14} for masking satellite galaxies. Credit: \citet{presotto14}.}
\label{fig:presotto}
\end{figure}

Despite the common use of this method, it suffers from two non-negligible problems: (1) it does not account for the ICL that overlaps with the BCG in the transition between the two components; (2) the ICL is contaminated by
the contribution of large galaxies in the cluster (see, e.g., \cite{presotto14}). \citet{presotto14} developed a method to obtain refined versions of typical BCG + ICL maps that can be obtained with simple surface brightness cuts. Their method focused mainly on the removal from the map of the light coming from satellite galaxies (the so-called {\em masking}). In Figure \ref{fig:presotto}, a comparison is shown between the standard method of surface brightness cut (blue lines and symbols) and the results with their approach (red lines and symbols). They find that the standard surface brightness cut method systematically overpredicts the fraction of ICL as a function of distance from the
center, independently of the particular cut used. It must be noted, however, that there are observational studies (e.g., \cite{montes21} for the Abell 85 cluster) that found the opposite trend, i.e., the surface brightness cut method gives a lower fraction of ICL than, e.g., the profile fitting method. Figure \ref{fig:presotto} shows also that a standard surface brightness cut has a steep increase from the core to around 100 kpc (in the particular case of MACS J1206.2-0847, which is a massive galaxy cluster at $z\sim 0.4$ and part of the CLASH sample \cite{postman12}) followed by a plateau. On the other hand, Presotto et al.'s masking causes the ICL contribution to drop at large radii, suggesting that most of the ICL is concentrated close to the BCG, which is in good agreement with several recent observational and theoretical results (e.g., \cite{kravtsov18,demaio20,contini20,contini21}).

\begin{figure}[H]
\includegraphics[width=13 cm]{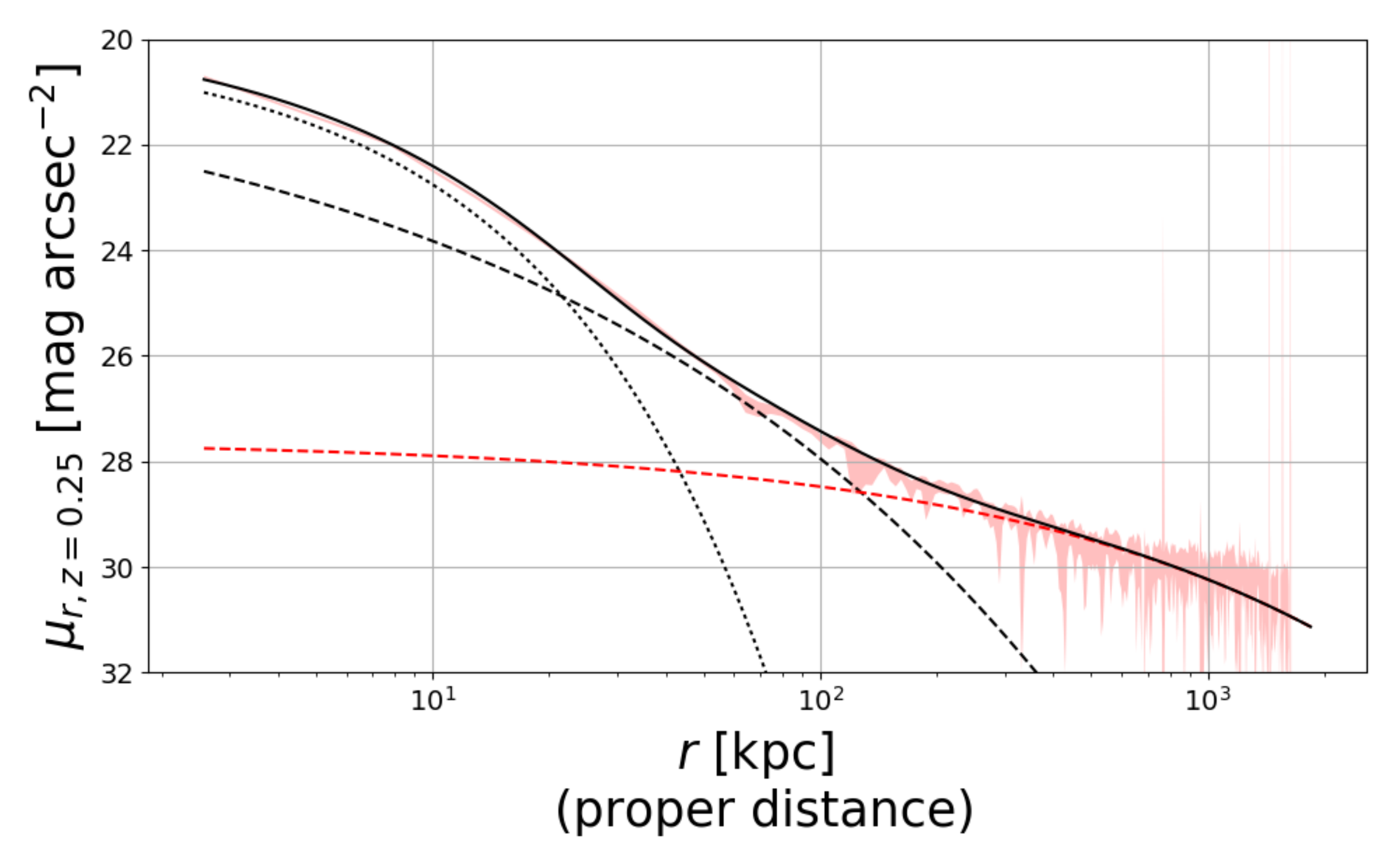}
\caption{The BCG + ICL light profile from \citet{zhang19} resulted from the stacking of around 300 clusters at redshift $0.2<z<0.3$. The authors found that it can be approximated with three Sersic components (black solid line): a core (dotted line), a bulge (dashed line) and diffuse (red dashed line) components. See text for further details. Credit: \citet{zhang19}.}
\label{fig:zhang19}
\end{figure}

The other most common method consists of using functional forms, such as a double/triple Sersic profile, to fit the light distribution (see references above). There are several ways to achieve it. For example, \citet{montes21} used the code GALFIT (\cite{peng02})  to map the 2D distribution of each component, BCG and ICL, with a double Sersic profile (one for each component). \citet{zhang19} used a triple Sersic profile to 1D fit the azimuthally averaged surface brightness stacked profile of 300 BCG + ICL systems. They found that, as shown in Figure \ref{fig:zhang19}, the overall profile can be approximated with the sum of a core, a bulge and a diffuse components (a similar conclusion has been reached earlier by \citet{kravtsov18}). The three components are dominant at different distances from the center. According to the parameter of the fit by \citet{zhang19}, the core is dominant within 10 kpc, the bulge is dominant between 30 and 100 kpc, and the diffuse component is dominant outside \mbox{200 kpc}. An advantage of this method lies in the fact that it can separate the two components in a more reliable way than a surface brightness cut where BCG and ICL overlaps, but it is strongly dependent on the functional forms chosen. For instance, \citet{zibetti05} showed that the distribution of the ICL can be described with an NFW profile \cite{navarro97}. The idea to link the ICL distribution with that of the DM has been used also in theoretical studies such as \cite{pillepich18,contini20,contini21}. In \citet{contini21}, the BCG + ICL mass distribution is described by the sum of three different profiles: a Jaffe \cite{jaffe83} profile for the bulge, an exponential disk and a modified version of an NFW profile for the ICL. I will come back on this topic in \mbox{Section \ref{sec:tracer}.}

It is worth mentioning another approach for detecting the ICL that has recently been having some success, a method that makes use of multiscale, wavelet-based algorithms. There are several examples of this approach that have been used in recent years (\mbox{e.g., \cite{darocha08,guennou12,adami13,ellien19,ellien21}}). One of the latest, just to quote an example, is the code called \mbox{DAWIS (\cite{ellien21}} and references therein). DAWIS is an algorithm, based on wavelet representation, built to restore the unmasked light distribution of given sources as much as possible. \mbox{\citet{ellien21}} compared the performance of DAWIS with the more common methods described above and found that it can separate the ICL from other sources more efficiently (in the sense that DAWIS performs better) than other methods and is also able to recover a larger quantity of ICL given the way it treats the sky background noise. For readers interested in the details of DAWIS and similar former algorithms of the same family, I refer them to \citet{ellien21} and references therein.

\subsection{Theoretical Methods}\label{sec:theo_meth}
Given the fact that, by definition, the ICL component is made of stars that are not gravitationally bound to any galaxy in the cluster, but only to the cluster potential, the natural way to define it would be to find some binding condition such that, all the stars obeying to it can be classified as ICL stars. This condition can be given by the {\em binding energy} of star particles with respect to the cluster galaxies. Without entering the details of the method, it is possible to calculate the gravitational potential energy as a function of radius of any given galaxy. In this way, one is able to measure the binding energy of each star particle to each galaxy and collect all those stars that are not bound to any galaxy (see \mbox{e.g., \cite{murante04,sommer-larsen05,murante07,dolag10,rudick11}}). The collection of these stars will constitute the ICL component of the cluster.

\begin{figure}[H]%
\includegraphics[width=13 cm]{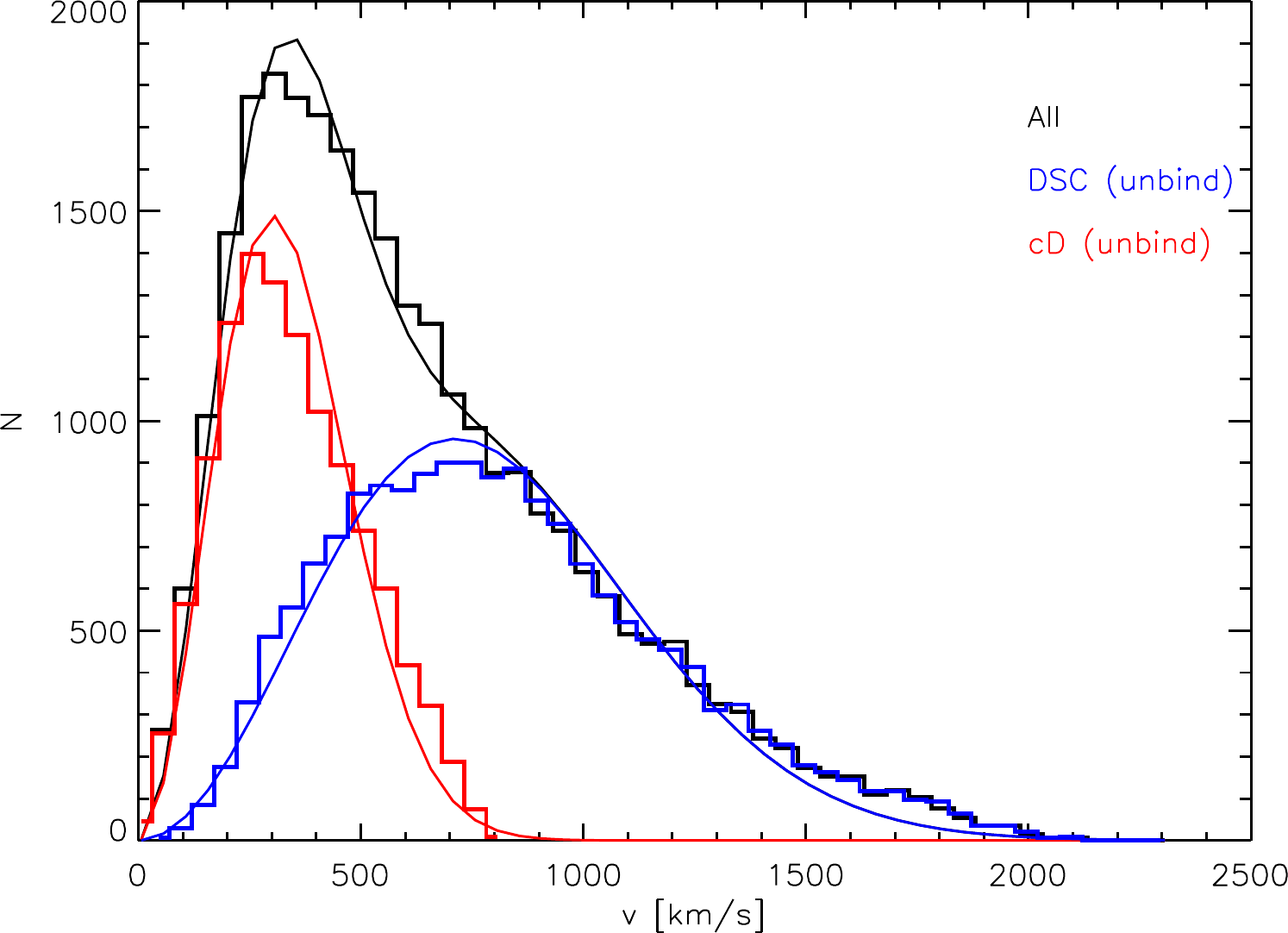}
\caption{Total velocity distribution of BCG + ICL is marked with a black histogram and its double Maxwellian fit with a grey line. The red and the blue histograms show the velocity distributions of the BCG  (called cD in the plot) and ICL (called DSC, diffuse light component, in the plot) stars, and the corresponding red and blue lines are the two separated Maxwellian distributions. Credit: \mbox{\citet{dolag10}}.}
\label{fig:dolag}
\end{figure}

However, despite the binding energy method being efficient in finding stars not bound to satellite galaxies, it does not succeed in separating the BCG from the ICL. Indeed, given the fact that the BCG is placed at the center of the cluster potential, it is not possible to distinguish its mass density profile from that of the cluster itself. In order to completely separate the two components, a few accompanying solutions have been suggested. The most common one has been introduced for the first time by \citet{dolag10}, who used the kinematics of the two components to separate the BCG from the ICL. They found that it is possible to fit the velocity distribution of BCG + ICL with two Maxwellians having different velocity dispersions, and suggested that they correspond to the two components. Figure \ref{fig:dolag} from \citet{dolag10} shows the result of their fit: The total velocity distribution of BCG + ICL is marked with a black histogram and its double Maxwellian fit with a grey line. The red and the blue histograms show, instead, the velocity distributions of the BCG and ICL stars, and the corresponding red and blue lines are the two separated Maxwelllians.

Another method usually used in numerical simulations (see, \cite{rudick09,rudick11} and others) relies on the {\em three-dimensional mass density}. This method consists of calculating the mass density of each star particle within a sphere of radius equal to the distance of a given {\em N}-th nearest neighbor. The further assumption is a threshold density below which the particles can be assigned to the ICL component. The weaknesses of this method lie mainly in identifying ICL particles in high-density regions, and especially in the cluster center. A way to partly overcome this problem has been proposed by \citet{rudick11}, where they look at the density history of each particle, rather than that at a given time, and a particle already classified as ICL remain classified as ICL regardless of its future evolution. Clearly, with respect to the aforementioned method, a density-based approach introduces more free parameters, depending on the level of accuracy that one wants to achieve.

In the list of numerical methods, it is also worth mentioning  some semi-analytical approaches. In semi-analytic models, the definition of the ICL does not constitute an issue, given the fact that the amount of ICL is provided by the solution of a set of equations, and its properties depend only on the particular implementation used to describe its formation and evolution. There have been several attempts to describe the formation of the ICL in semi-analytic models, but for the sake of simplicity, I will report only two amongst the most recent.

\citet{guo11} assumed that the ICL forms from the stellar component of satellite galaxies that are subject to tidal forces after their parent substructures have been totally stripped. These kinds of galaxies are usually referred to as {\em orphans} to indicate that their parent subhalo went under the resolution of the simulation. At the pericenter, the main halo density is compared with the average baryon density of the satellite within its half mass radius. If the former is larger than the latter, the satellite is assumed to be disrupted and its stars are assigned to the ICL component. This approach, however, suffers from important limitations. As partly mentioned in Section \ref{sec:intro}, a complete disruption of satellites is less likely than a partial stripping, i.e., some amount of mass stripped. It is indeed more likely that satellite galaxies are subject to several stripping events rather than being totally disrupted in a
single one. Moreover, it has been shown that the stripping of the stellar component starts before the complete stripping of the DM subhalo (see, e.g., \cite{villalobos12,smith16}).

A more realistic representation of the stellar stripping was implemented first in \citet{contini14}, then revisited in \citet{contini18,contini19}. In the so-called {\em tidal radius model}, the authors assumed that a satellite can lose
stellar mass in a continuous way, with several stripping events. The model, at each time step, calculates the tidal radius $R_t$ of the interaction between the cluster potential and the satellite, at the distance of the satellite from the
cluster center. A satellite is modeled as a two-component system, bulge and disk. If the tidal radius $R_t$ is smaller than the radius of the bulge, the satellite is assumed to be destroyed, but if it is larger than the bulge radius and smaller than the radius of the satellite, the mass of the disk in the shell between the two radii is stripped and ends up as the ICL component. 
This method is not only applied to orphan galaxies but also to satellites that still have a DM subhalo. The extra requirement for these satellites is that the half mass radius of the subhalo is smaller than the half mass radius of the disk, which translates into a substantial amount of DM stripped (in accordance with numerical simulations). To account for the stellar mass that gets unbound during galaxy mergers (see, e.g., \cite{conroy07,murante07,burke12,oliva14,zhang16} and others), the model also considers  the {\em merger channel}. At each merger, minor or major, 20\% of the stellar mass of the satellite that is merging with the central galaxy is added to the ICL component. I will return to the importance of both channels for the formation of the ICL in Sections \ref{sec:formation} and \ref{sec:properties}.

\section{Formation Mechanisms}\label{sec:formation}
In Section \ref{sec:intro}, I have briefly listed the most important physical mechanisms that have been suggested in order to explain the formation and evolution of the ICL. Some of them have been ruled out ({in situ} star formation and disruption of dwarf galaxies) by several authors in the past years; therefore, in this section I will cover the details of the remaining three important channels: pre-processing, mergers and stripping of galaxies. {A caveat
that is worth mentioning for the following discussion concerns the definitions of the CL coming from any of those processes. Indeed, even with numerical simulations it is not possible, for example, to separate in a clear way stellar stripping from mergers in the vicinity of the cluster center. The results will always depend on the particular definitions used. However, regardless of the definitions used to separate the different channels, once the stars are removed from satellites, they will be added to the ICL component, which only considers  the DM potential well of the cluster.}


\subsection{Pre-Processing}\label{sec:pre-proc}

What are we referring to when we use the term {\em pre-processing}? Usually, the term pre-processing refers to any material that ended up somewhere but that has been processed elsewhere. The ICL is a clear example of such material. Indeed, part of it has been pre-processed in galaxy groups and then later accreted in clusters. This is a natural consequence of the hierarchical formation of structures. DM haloes continue accreting smaller objects with time, thus increasing their mass. In a very simplified manner, the process can be thought of as follows: at high redshift, DM haloes tend to be smaller and less massive than present day clusters, but galaxies within them start to interact with each other and eventually merge. This lead to the formation of the first ICL stars in galaxy groups (usually called IGL). However, due to the hierarchical nature of the formation of DM haloes, these high redshift
smaller haloes will be later accreted by other larger objects, carrying the ICL already formed. In summary, we refer to pre-processed ICL any ICL formed in the past and later accreted ({With the term
{\em accretion}, I mean that the ICL is added to the entire cluster, and becomes part of the ICL already present.}). When galaxies were originally centrals, they probably had some ICL associated with them. By being accreted (i.e., they become satellites), they carry their ICL with them, but due to the tidal interactions with the potential well of the new DM halo, their ICL is stripped and becomes part of the diffuse light already present in the accreting halo.

\begin{figure}[H]
\includegraphics[width=10 cm]{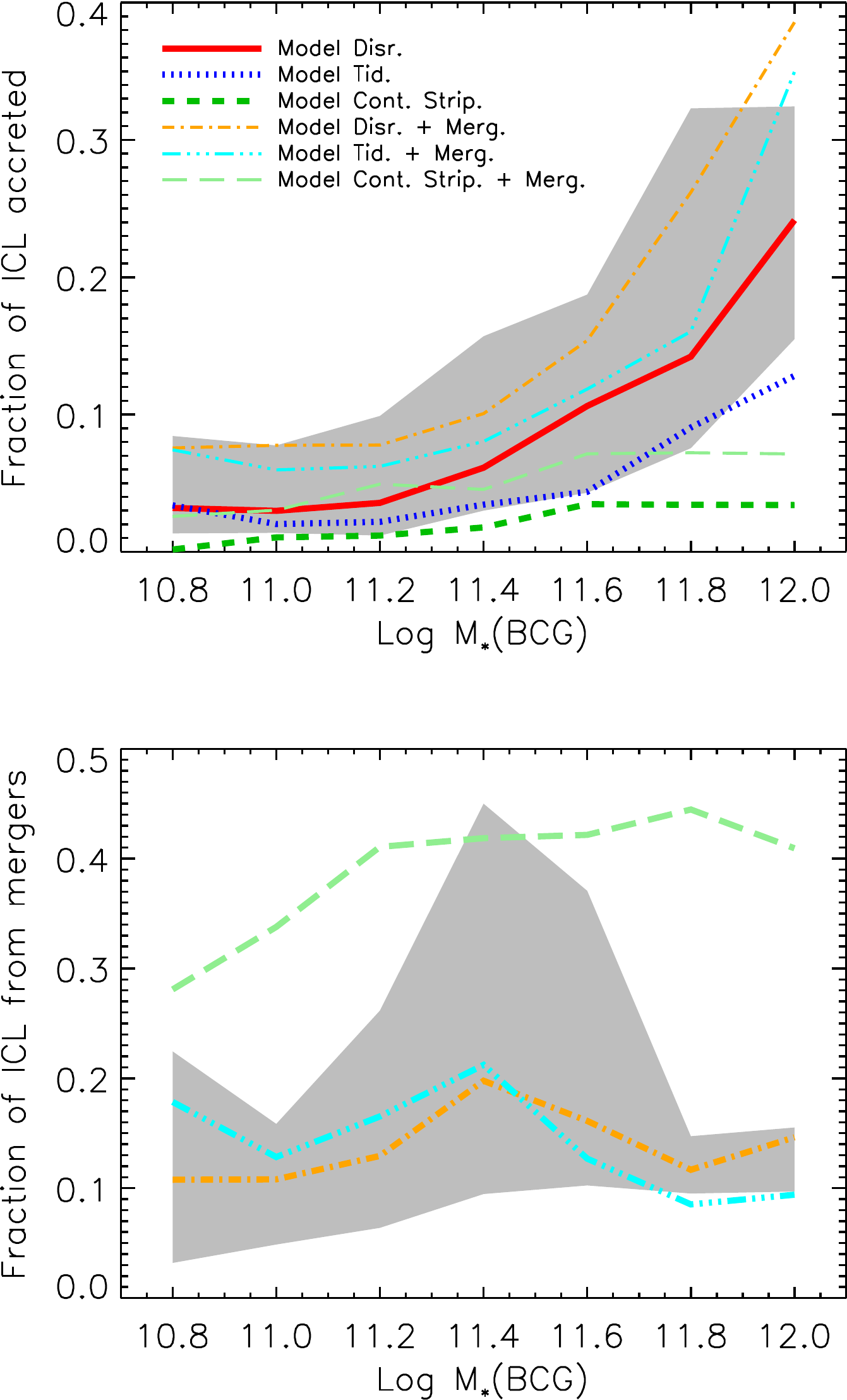}
\caption{{Top} panel: fraction of ICL that has been accreted as a function of the mass of the BCG for different flavors of the model described by \citet{contini14}. {Bottom} panel: fraction of the ICL coming from the merger channel as a function of BCG stellar mass. Credit: \citet{contini14}.}
\label{fig5}
\end{figure}

In terms of DM, pre-processing has been shown to be quite an important process. For example, \citet{han18} found that almost 50\% of members in present day clusters were satellites of other hosts, although the fraction depends on the particular mass accretion history of each cluster. Galaxies sit inside DM haloes, so it is natural to expect that a given fraction of the current ICL was produced in the past and then accreted. \mbox{\citet{contini14}} showed that pre-processing is particularly important for high mass BCGs that reside in clusters, and it can contribute to 30\% of the total amount of ICL for the largest BCGs and slightly less than 10\% for the smallest BCGs (which are called BGGs when residing in groups), as shown in the top panel of Figure \ref{fig5}.

A direct example of an ongoing pre-process mechanism is given by the Virgo cluster. Virgo is a nearby galaxy cluster with a moderate richness, and given its vicinity, its structure has been well studied. Precisely its structure makes Virgo an interesting object to study. Virgo can be divided in two subclusters: One of them is centered around M87, a giant elliptical, and the other is centered around another giant elliptical, M49. There are other smaller subclusters that can be identified, such as those surrounding the other two elliptical galaxies, M59 and M60, and a few clouds of galaxies named M, W and W$'$. Clearly, Virgo is continuing to assemble today and it is far from being at the final stage of its assembly. This makes it particularly interesting for studies that look at the connection between the dynamical state of clusters and their ICL amount and distribution. \citet{mihos17} recently showed that the W$'$ cloud in Virgo is a nice example of pre-processing played by groups of galaxies. Indeed, the ICL already formed in the W$'$ cloud will soon be accreted in the main body of Virgo, and given the strong tidal fields, it will be stripped and dispersed in the global ICL already present.

\subsection{Mergers}\label{sec:mergers}

Another possible mechanism for forming diffuse light is given by galaxy mergers, due to the relaxation processes that take place during the very moment of a merger. Before going into the detail of this channel, it is worth mentioning a caveat. Mergers are not a well-defined process, in the sense that there is not a clear definition of when a merger starts. Of the possible definitions that can be found, all of them would be affected by the fact that part of the stars that will become unbound can actually be classified as belonging to the stripping channel (see discussion in \cite{contini18}). However, in the final stage of a merger, a given fraction of the stars belonging to the satellite galaxy merging with the central can become unbound and disperse in the ICL component of the cluster.

The literature has plenty of attempts to theoretically model this channel. In hydrosimulations, it is pretty straightforward since it depends on the physical interactions between particles. What is not clear yet, is exactly how to define the moment of when the merger starts, so that particles coming from this channel can be separated from those stripped during the process. Conversely, in semi-analytic models, the definition of a merger is very neat (but it does not necessarily mean the right one), and a merger is defined as the moment when the dynamical friction time of the merging satellite is zero.

\citet{murante07} focused on the formation of the ICL by means of hydrosimulations and found that most of it (75\%) forms through mergers between either the BCG or other massive galaxies, leaving just a small percentage to the stripping channel. On the other hand, \citet{contini14} showed that the merger channel contributes to the total ICL with just 15\%, as shown in the bottom panel of Figure \ref{fig5} 
(cyan line), and the rest given by stellar stripping. In \citet{contini18}, the authors argued that this huge difference can be easily explained by the two different definitions of the merger channel used. Indeed, they found that 70\% of the ICL coming from stellar stripping is produced in the innermost 100 kpc, a relatively small region within which a satellite can be considered in the process of merging. If they included this part of ICL to the merger channel, they could reach 75\%, the same percentage found by \citet{murante07}. As mentioned in Section \ref{sec:theo_meth}, in \citet{contini14}, the authors assumed that 20\% of the mass of the satellite galaxy merging with the central ends up in the ICL component. This fraction comes from controlled simulations performed by \citet{villalobos12}, although in reality the fraction can be significantly different from case to case, possibly depending on the orbital circularity, satellite and BCG/halo mass and other important parameters. Similar implementations have also been used  in other semi-analytic models (see, e.g., \cite{monaco06,somerville08}).

From the observational point of view, understanding the relative contribution given by mergers to the ICL (and the same for stellar stripping) is far from being easy, simply because it is not possible to trace the past of the ICL stars. However, there are indirect methods that can provide an idea of how much mergers can contribute, and others by looking at the properties of the ICL (which I will discuss in Section \ref{sec:properties}). For instance, some studies (e.g., \cite{burke15,groenewald17}) pointed out the discrepancy between the observed and predicted stellar mass function, and argued that a possible way to significantly reconcile the two is by invoking a significant mass loss during galaxy mergers, quantified to be around 50\%  ({other theoretical
studies, e.g., \citet{contini17a,contini17b}, have shown that the observed and predicted stellar mass functions can be matched by implementing both stellar stripping and mergers (assuming a much lower percentage than 50\%).}). However, such a high fraction of mass that ends up in the ICL component rather than the BCG seems to be in contrast with the observed properties of the ICL (see Section \ref{sec:properties}).

\subsection{Stellar Stripping}\label{sec:stripping}

The last important channel for the formation of the ICL is the stellar stripping from galaxies orbiting around the center of the cluster. Stellar stripping applies to all galaxies within a cluster, but tidal forces are responsible for it becoming stronger in the innermost regions close to the cluster center. The efficiency of the stellar stripping, i.e., how strong the interaction between a central galaxy in a halo and a satellite is, depends mainly on two factors: The environment, meaning the mass of the halo and the distance from the center, and the mass of the satellite. The dependence on the latter comes from dynamical arguments. Indeed, due to the dynamical friction \cite{chandrasekhar43} that subhaloes (and so galaxies within them) experience when accreted in larger haloes, they orbit around the potential well of the halo by feeling a drag force, caused by the surrounding mass distribution, which tend to slow them down by losing kinetic energy and momentum. This drag force is directly proportional to that mass of satellites, i.e., the biggest travel faster to the center, where they are more likely to be subject to stripping than less massive satellites (see \mbox{also \cite{contini12,roberts15,contini15,kim20}}).

The other important factor for the efficiency of stellar stripping is the environment in which satellites are, and this means the mass of the host halo and the distance from the center. More massive haloes are less concentrated than less massive ones, i.e., groups are more concentrated than clusters (e.g., \cite{gao11,contini12,prada12}). Therefore, the efficiency of stripping is expected to be higher in groups rather than clusters, at least in the innermost regions. However, the tidal radius, that is, the radius at which tidal forces are stronger than the gravity of the satellite, is inversely proportional to the mass of the halo and directly proportional to the mass of the satellite and its distance from the center \cite{binney08}. This means that, a satellite with the same mass and at the same distance in a group or cluster will experience a larger stripping if it resides in a cluster.

\begin{figure}[H]
\includegraphics[width=13.5 cm]{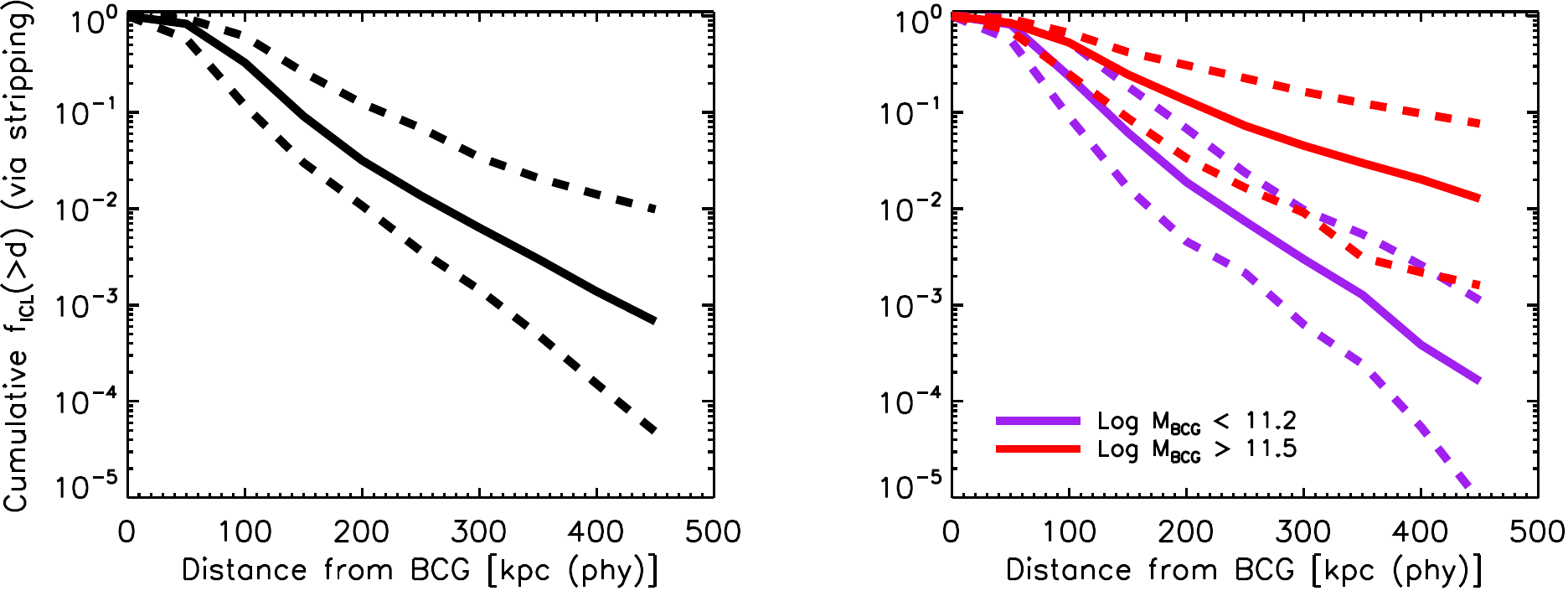}
\caption{{Left} panel:
cumulative fraction of stellar mass in ICL produced via stellar stripping as a function of distance from the BCG. {Right} panel: same as left panel, but for BCGs with $\log M_* [M_{\odot}] <11.2$ (purple lines) and BCGs with $\log M_* [M_{\odot}] >11.5$ (red lines). In both panels, solid and dashed lines represent the median, the 16th and 84th percentiles of the distributions, respectively. Credit: \citet{contini18}.}
\label{fig:contini18a}
\end{figure}

A clear example of the efficiency of stellar stripping in the innermost regions of groups and clusters is given in Figure \ref{fig:contini18a} (from \citet{contini18}). The left panel shows the cumulative fraction of mass in ICL that has been produced by stellar stripping as a function of distance from the BCG. The solid black line, which represents the median of the distribution, shows that only 10\% of the total ICL coming from stellar stripping is built-up at distances farther than 150 kpc, which means that almost all the ICL produced via this channel actually comes from stripping events that happened in the innermost 150 kpc. The right panel of Figure \ref{fig:contini18a} shows the same information shown in the left panel but for two samples of BCGs: less massive than $\log M_* [M_{\odot}] <11.2$ (purple lines), and more massive than $\log M_* [M_{\odot}] >11.5$ (red lines). The panel clearly shows that for more massive BCGs, most of the ICL coming from stellar stripping tends to be produced at larger distances than for less massive BCGs. This result is explained by the trend discussed above, i.e., less massive BCGs are likely to reside in less massive haloes, and these objects are more centrally concentrated than their more massive counterparts.

\begin{figure}[H]
\includegraphics[width=13.5 cm]{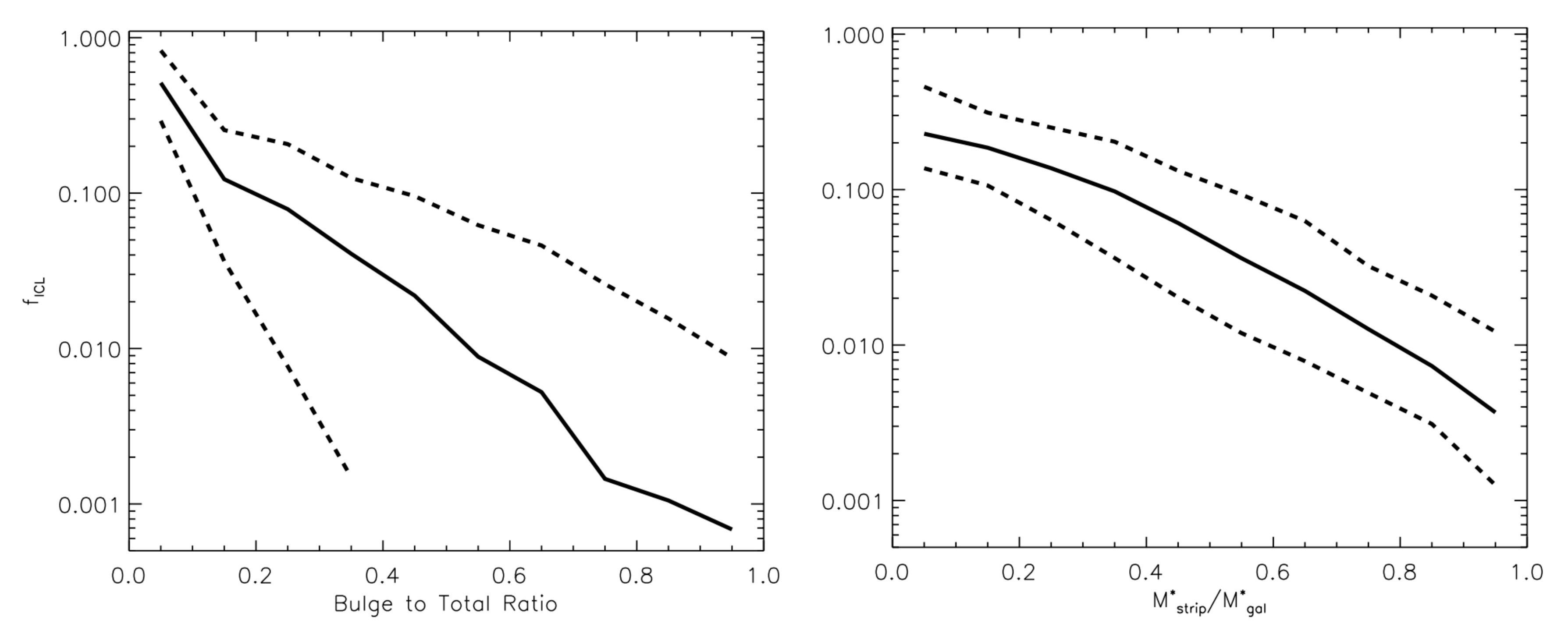}
\caption{{Left} panel: fraction of ICL mass via stellar stripping as a function of the bulge-to-total mass ratio of galaxies subject to stripping events, within 100 kpc from the halo center. {Right} panel: same information as the left panel but as a function of the ratio between the amount of mass stripped and the mass of the satellite at the moment of stripping. In both panels, solid and dashed lines represent the median, the 16th and 84th percentiles of the distributions, respectively. Credit: \citet{contini18}.}
\label{fig:contini18b}
\end{figure}

Stellar stripping is stronger in the innermost region of the haloes, and massive satellites (due to dynamical friction) reach that region faster. It is, then, reasonable to expect that they are the most important contributors to the ICL. This prediction was made in \mbox{\citet{contini14}}, and later confirmed by several observations (e.g., \cite{montes14,demaio15,demaio18,montes18,montes21}). In \citet{contini18}, they also looked at what kinds of galaxies contribute the most to the production of ICL via stellar stripping. Figure \ref{fig:contini18b} (from \citet{contini18}) shows the amount of ICL produced in the innermost 100 kpc via stellar stripping, as a function of the bulge-to-total ratio (left panel), and the amount of mass stripped over the mass of the satellite at the moment of stripping (right panel), for satellite galaxies involved in stripping events. The trend appears pretty clear: disk-like galaxies (B/T < 0.4) contribute the most to the production of ICL, while ellipticals and spheroidals contribute in a marginal way (for details see \cite{contini18}). The message shown in the right panel is as interesting as the previous one. Indeed, most of the ICL produced by stellar stripping in the innermost 100 kpc comes from small/intermediate stripping events that cut just <30\% of the mass of the satellite at the moment of stripping. The total disruption of the satellite is a very unlikely event.

All the above theoretical predictions have important consequences on the properties of the ICL. If intermediate/massive satellites, that are likely to be disk-galaxies, contribute the most to the production of ICL via small/intermediate stripping events, it is reasonable to expect that properties of the ICL, such as colors and metallicity, assume given values that are different from those that one would expect if the ICL is mainly produced via mergers or disruption of dwarf galaxies. This is exactly the approach followed by observational studies (e.g., \cite{montes14,demaio15,morishita17,iodice17,montes18,zhang19,montes21}): by looking at the typical colors and metallicity of the ICL, it is qualitatively possible to figure out the process responsible for the formation of the ICL, and the major contributor to it. In addition, by comparing the typical properties of the ICL with those of the BCG, it is possible to learn more about their  history. 
I will address these points in the next section.

\section{Properties of the ICL}\label{sec:properties}

The properties of the ICL are important for understanding which, among the aforementioned mechanisms, is responsible for the formation and evolution of the diffuse component in groups and clusters. In the following, I will summarize the latest achievements in studying the most important properties, i.e., the ICL fraction, the ICL colors, metallicity and age. These properties are not only  important in the context of the physical mechanisms, which led to the formation of the ICL, but can give information on the dynamical state of the cluster where it resides.

\subsection{ICL Fraction}\label{sec:iclfrac}

Keeping in mind the issues regarding the definition of the ICL and the different methods to separate it from the rest of the light discussed above, the easiest quantity to study is likely the ICL fraction (hereafter $f_{ICL}$), that is, the ratio between the amount of stellar mass in ICL and the total stellar mass within the virial radius of a cluster. Why is $f_{ICL}$ so important? From a theoretical point of view, we would expect to have higher fractions of ICL in those objects that are more dynamically evolved (\cite{murante07,rudick11,martel12,contini14} and others), because in these kinds of objects, the probability of stellar stripping, as well as mergers between galaxies, is higher.

There is no general agreement of a possible dependence of $f_{ICL}$ on cluster mass among different studies, both observational and theoretical (\cite{murante04,murante07,cui14,contini14,alamo17,mihos17,ko18,spavone18,montes18,tang18,jimenez18,jimenez19,henden19,zhang19,raj20,spavone20,canas20,ellien21,ragusa21,furnell21,kluge21} just to quote some theoretical and the most recent observational studies). Some theoretical works (e.g., \cite{lin04,purcell07,murante07,henden19}) find an increasing $f_{ICL}$ with cluster mass, others (e.g., \cite{dolag10,henriques10,contini14}) no relation, and others (e.g., \cite{cui14}) a slight decreasing $f_{ICL}$ with increasing cluster mass. Overall, by collecting all the results so far obtained and considering the scatter, $f_{ICL}$ is found to be independent on halo mass, and range between $\sim$5\% (group scale, lower limit) and $\sim$50\% (cluster scale, upper limit) at the present day ({I refer
those readers interested in a collection of the most recent $f_{ICL}$ derived in observed and simulated objects to \mbox{\citet{tang18,kluge21,ragusa21,furnell21}}}), with a quite large scatter over the whole halo mass range probed. An important caveat here must be noted: $f_{ICL}$ strongly depends on the radius within which it is measured, which is usually different from study to study. Then, attention must be paid when comparing fractions obtained by different authors.

\begin{figure}[H]
\includegraphics[width=13.5 cm]{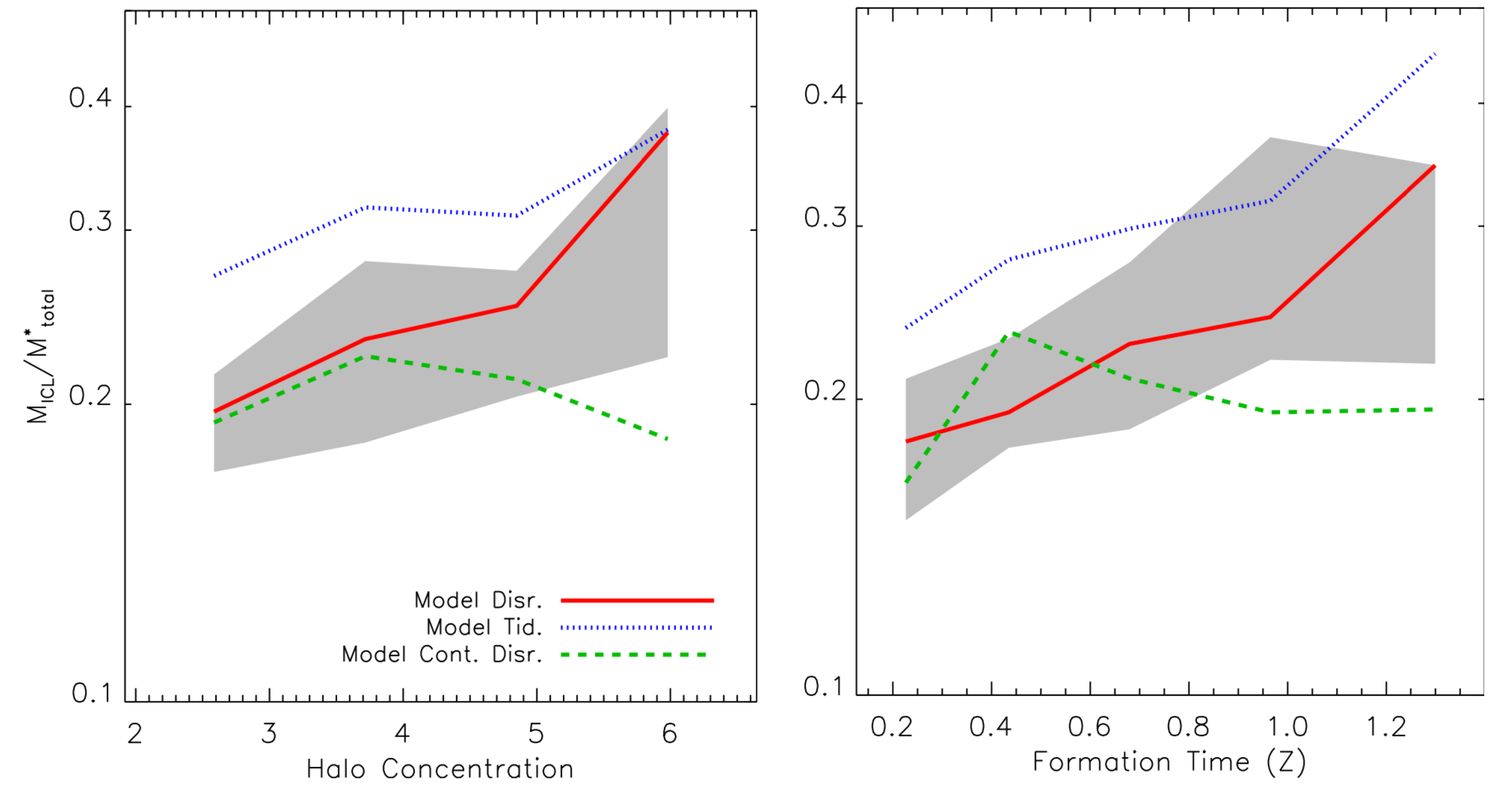}
\caption{{Left} panel: fraction of ICL as a function of the concentration (defined as the ratio between the scale radius by assuming an NFW profile and the virial radius $R_{200}$) of 53 haloes with mass $M_{200} [M_{\odot}/h] > 10^{14}$. {Right} panel: same information but as a function of the halo formation time, defined as the time the first main progenitor acquired half of its final mass. The gray shaded regions show the 20th and 80th percentiles of the distributions. Credit: \citet{contini14}.}
\label{fig:contini14a}
\end{figure}

The scatter in the $f_{ICL}-M_{halo}$ relation can be explained, mostly, in terms of the concentration and formation time of the particular halo in which the ICL formed. \mbox{Figure \ref{fig:contini14a}}, taken from \citet{contini14} (set of implementations of the ICL in a semi-analytic model), shows that $f_{ICL}$ depends on both concentration and halo formation time and increases for more concentrated and old haloes (except for a model for which the reader can find the details in \citet{contini14}), which is a direct consequence of the dynamical evolutionary state of the haloes. In more concentrated haloes, the tidal forces are stronger and the probability of stellar stripping increases. Similarly, older haloes have more time to produce ICL or acquire it from smaller haloes. The concentration and the halo formation time can explain the large scatter that is observed in the $f_{ICL}-M_{halo}$ relation for high mass haloes, but, once we plot the same information for smaller haloes, on group scale, the correlation disappears. In \citet{contini14}, the authors argued, and proved, that in this low halo mass regime, the scatter in the $f_{ICL}-M_{halo}$ relation is driven by the number of intermediate/massive satellites which, as argued above, are those that contribute the most to the production of the ICL (I will come back to this topic in Section \ref{sec:colmet}). The scatter in the accretion of intermediate/massive satellites drives the scatter in the $f_{ICL}-M_{halo}$ relation in a way that the larger the number of intermediate/massive satellites accreted, the higher the fraction of ICL. It must be noted, however, that the above explanation concerns the scatter in the $f_{ICL}-M_{halo}$ relation predicted by theoretical models, and the same exercise can be performed in hydrosimulations, but the {\em observed} scatter might depend also on other factors, such as sample selection and data quality, as well as the definitions used by different authors to separate the ICL from the rest.

An interesting aspect related to $f_{ICL}$ is given by its evolution with redshift, i.e., how it depends on time. For the reason discussed above, if the ICL formation is linked to processes such as stellar stripping and mergers (as it is), from a theoretical point of view, we have to expect $f_{ICL}$ to grow with cosmic time, simply because the growth of the ICL would be linked to the growth of the clusters themselves. While almost all theoretical models/simulations agree on the fact that the bulk of the ICL forms late ($z<1$), and that $f_{ICL}$ grows with time, from the observational side there is no general consensus. Not only there are examples of significant quantities of ICL in high-z clusters, e.g., \citet{ko18}, but recent observational results either show a mild or no correlation between $f_{ICL}$ and time (e.g., \cite{guennou12,montes18}), or even a substantial evolution after $z\sim 0.5$ (e.g., \cite{burke15,furnell21}). Once again, the inconsistency of the observational results can be attributed in part, if not the most, to the intrinsic problem of isolating the ICL (different definitions), which becomes even more serious at higher redshift due to its faintness and the effect of surface brightness dimming.

\begin{figure}[H]
\includegraphics[width=10. cm]{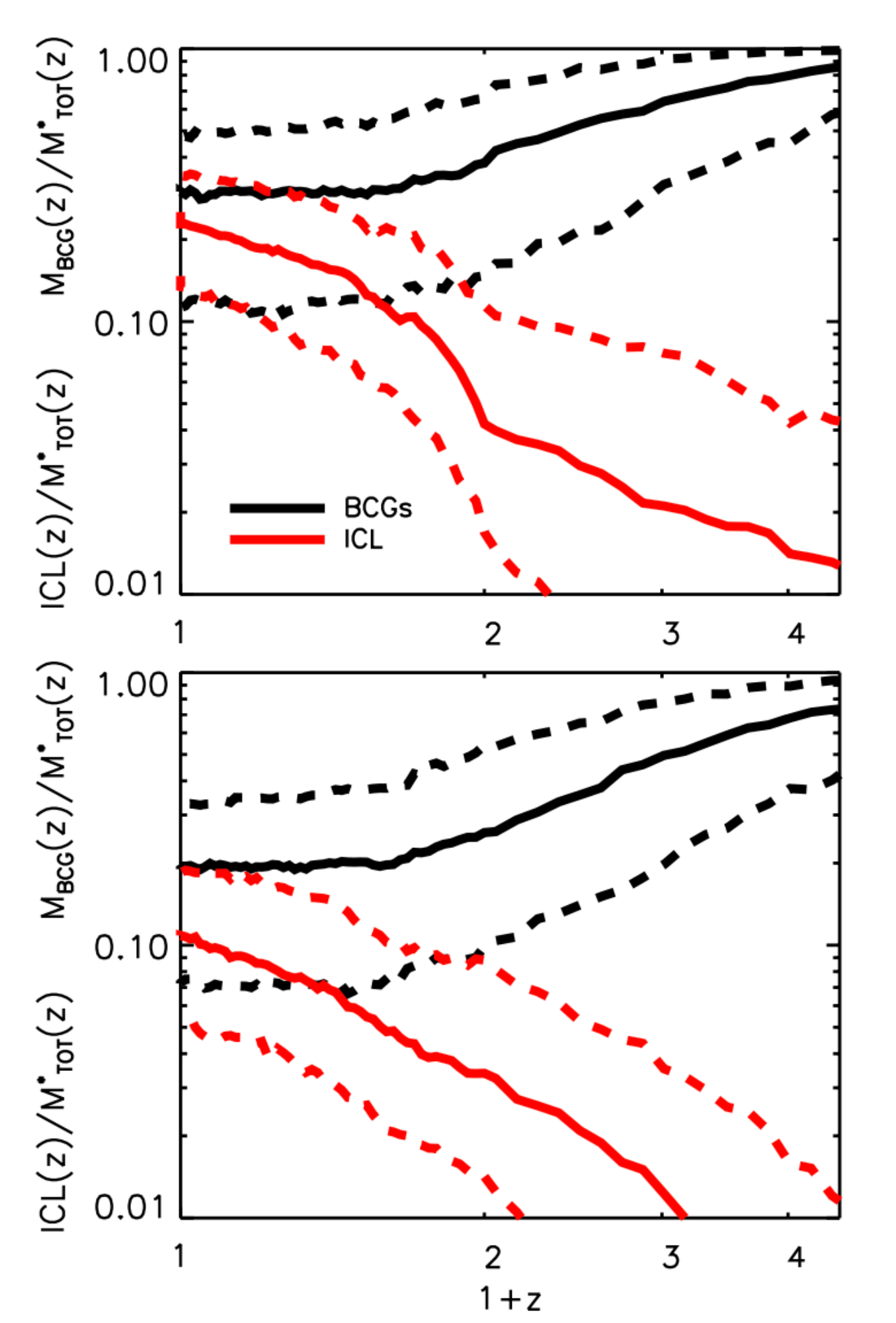}
\caption{ICL (red lines) and BCG (black lines) fractions as a function of redshift, for a model in which the ICL forms via stellar stripping and mergers ({top} panel, see description of the model in Section \ref{sec:theo_meth}), and
a model where the ICL forms only via mergers ({bottom} panel). Solid lines represent the median, while dashed lines represent the 14th and 16th percentiles of the distributions.  Both fractions have been calculated within $R_{200}$ of the halo at any given redshift. Credit: \citet{contini18}.}
\label{fig:contini18c}
\end{figure}

However, nowadays, a scenario is commonly accepted  where the BCG and the ICL co-evolve (to which I will come back to in Section \ref{sec:coevol}), for which the period after $z\sim 1$ and the present day is the time for a rapid growth of the ICL but of a slow and little growth of the BCG. This is supported by several numerical models/simulations but also by observational evidence (e.g., \cite{burke15,montes14,demaio18,montes18,demaio20}). In order to give an idea of what numerical models predict in the redshift range $0<z<1$ for the growth of the BCG and ICL, Figure \ref{fig:contini18c} shows how the ICL/BCG fractions grow with time after $z<1$, for the reference model in \mbox{\citet{contini18}} (top panel) and described in Section \ref{sec:theo_meth}, and a model where the ICL forms only through mergers, with the mass of the satellite that is equally split between BCG and ICL during each merger (bottom panel). Both panels show an interesting trend: while the BCG grows slower than the overall growth of the stellar mass in the cluster (decreasing fraction), the ICL grows faster (increasing fraction). As mentioned above, these behaviors of BCGs and ICL have important consequences on the whole evolution of BCG-ICL systems, and they are strictly linked to the growth of clusters. I will fully address this point in \mbox{Section \ref{sec:coevol}.}

The fraction of ICL in groups and clusters can provide some important information, as we have just seen, but much information is given by other fundamental properties of the ICL: colors, metallicity and age. Indeed, by looking at these properties, it is possible to obtain a direct picture of the responsibility of different formation mechanisms and by the size/morphology of the galaxies involved in the formation of the ICL.

\subsection{ICL Colors, Metallicity and Age}\label{sec:colmet}

ICL colors, metallicity and age are fundamental properties that can characterize the ICL in two possible directions. By considering the whole BCG + ICL system, the gradient (if any) from the BCG dominated distance to the ICL dominated one of either one of these properties can lead to important conclusions regarding the mechanism responsible for the formation of the ICL, while the intrinsic values of them can give an idea of which kinds of galaxies contributed the most. These properties have been studied extensively during  recent years (\cite{toledo11,melnick12,contini14,montes14,demaio15,edwards16,mihos17,iodice17,morishita17,montes18,jimenez18,demaio18,ko18,spavone18,jimenez19,zhang19,contini19,edwards20,gu20,raj20,montes21,ragusa21} just to quote the most recent ones), and most of the studies are converging to the following scenario: generally, there is a negative gradient of colors, metallicity and age from the innermost parts (BCG dominated) to the outermost regions (ICL dominated) ({it is important
to point out that the age-metallicity degeneracy is a challenge for disentangling these two quantities from colors alone}).

This evidence has quite an important consequence, i.e., that the ICL results in bluer, more metal-poor and younger stars 
than the BCG. These results also shed some light on the possible mechanism for the formation of the ICL. Indeed, if the violent relation processes during mergers are the main responsible mechanisms, one would expect no gradient, given the fact that mergers potentially mix up the population of stars in both components. On the other hand, such a result would be perfect in a scenario where stellar stripping is the main responsible mechanism. In this case, the stripping of stars from satellite galaxies would furnish the ICL component with stars that have properties different from those of the BCGs. The kind of satellites that are subject to stripping will reflect the colors/metallicity/age of the stars in the ICL. A negative gradient is expected in the case that most of the stars come from the outskirts of intermediate/massive galaxies, being bluer and more metal-poor than the average stars in the BCG.

The most recent observations (e.g., \cite{montes14,demaio15,morishita17,iodice17,spavone18,montes18,zhang19,ragusa21,montes21}) have found a clear radial gradient that, as just argued, favors stellar stripping to mergers as the main mechanism for the formation of the ICL. In addition, the majority of these studies also agree  on the fact that the typical colors and metallicities of the ICL are similar to those of intermediate/massive galaxies, meaning that the scenario of the ICL formation predicted in \citet{contini14} and later improved in \citet{contini18,contini19}, i.e., the ICL forms mainly from stripping of intermediate/massive galaxies after $z\sim 1$, is a good representation of what happens in real clusters. Indeed, in \citet{contini14} the authors argued the importance of deriving metallicities/colors so that observational evidence could help in constraining the models. In this regard, it is worth providing a panoramic of the most recent observational results regarding colors/metallicity or age of the ICL (or BCG + ICL).

\begin{figure}[H]
\includegraphics[width=12. cm]{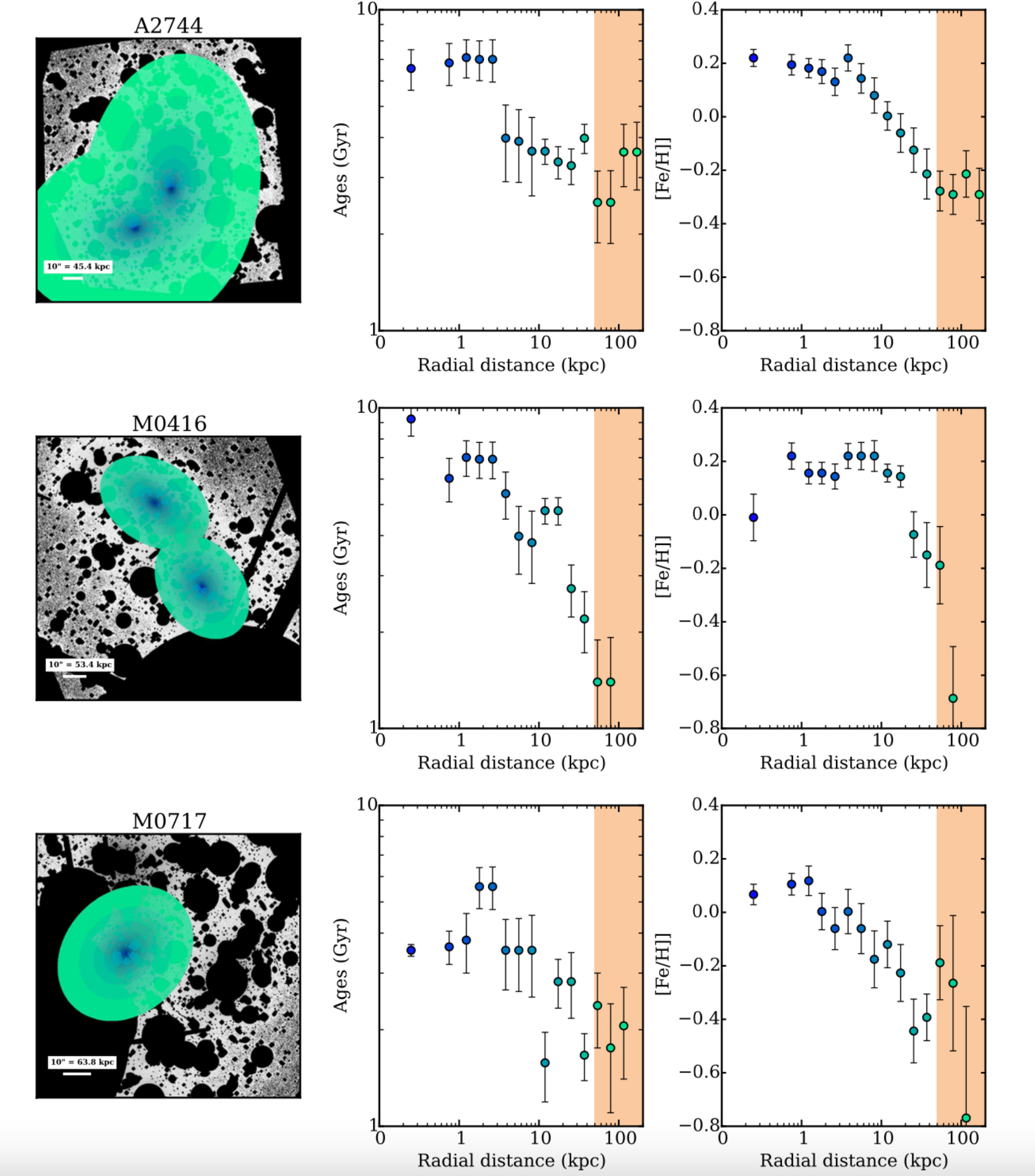}
\caption{A few examples of age and metallicity gradients for three clusters of the Hubble Frontiers field survey in the range $0.3<z<0.6$. The clear tendency for the stellar populations in these BCG + ICL systems is to become younger and more metal-poor with increasing distance from the center, i.e., in the regions dominated by the ICL. Metallicity and age are derived via stellar population synthesis models (see the original paper for the details of the modeling). Credit: \citet{montes18}.}
\label{fig:montes18}
\end{figure}

\citet{montes14} derived the stellar population properties of the ICL in the Abell 2744, which is a $z\sim 0.3$ massive cluster. The remaining frame colors of the ICL are bluer than the main galaxy members, and from the colors, the authors derived a mean metallicity of the ICL consistent with the solar one and an age $6\pm3$ Gyr younger than the mean age of the most massive galaxy members.  According to the data, the most plausible scenario for the formation of the ICL in A2744 is from the stripping of Milky Way-like galaxies with a similar metallicity, after $z\sim 1$. Recently
, \citet{demaio15} focused on four clusters at higher redshift, $0.4<z<0.6$, and found that for three of them, there is a metallicity gradient that goes from super-solar metallicities in the central regions dominated by the BCG to sub-solar in those dominated by the ICL. Moreover, they found that the ICL is mostly dominated by stars
coming from the stripping of galaxies with luminosity $L>0.2L^*$ (being $L^*$ the characteristic luminosity at the knee of the luminosity function). Combining all their results, they ruled out late mergers as a channel of the ICL formation in those clusters. Not much later, these results were confirmed by the same authors \cite{demaio18,montes18} with wider samples of clusters (see also the conclusions of \citet{jimenez18} for a similar wide sample of clusters, \citet{zhang19} for a much wider sample of clusters at similar redshifts, and \citet{jimenez19} and \citet{gu20} for the Coma cluster). \mbox{Figure \ref{fig:montes18}} shows the age and metallicity gradients of the BCG + ICL in three of the six clusters analyzed by \citet{montes18} (similar gradients are found in the other three).  Going from the innermost regions where the contribution is dominated by the BCGs, to the outer ones where, instead, the contribution is dominated by the ICL, the stellar populations tend to become younger and more metal-poor, with different values depending on the particular BCG + ICL system. ({{Colors, age and
metallicity have been derived, also taking advantage of CGs and PNe for nearby objects (see the introduction for references). For readers interested in knowing more on how CGs and PNe can be helpful in this context, I refer them to studies such as \citet{longobardi15,longobardi18} and references therein.}}).

\citet{morishita17}, with the same sample of six clusters from the Hubble Frontiers Field at $0.3<z<0.6$ analyzed by \citet{montes18}, constructed the radial color and stellar mass profiles of their BCG + ICL via SED fitting. Similarly to the studies quoted above, they found negative color gradients with increasing distance from the BCG, and an average stellar population age of the ICL of $\sim$2 Gyr probably from stripping of dwarf/intermediate ($\log M_* <10$) galaxies. \citet{edwards20} studied a sample of \mbox{23 local} BCG + ICL systems and determined their stellar population age and metallicity. They found that the ICL is best modeled with younger and less metal-rich stars than those in the BCGs and found evident gradients in both age and metallicity from the BCG to the ICL dominated regions. Their results suggest that the inner regions formed quickly and early, while the ICL formed more recently and continues to assemble via minor mergers. Very recently, \citet{montes21} investigated the ICL properties of the Abell 85 cluster taken with the Hyper Suprime-Cam on the Subaru Telescope. They measured the radial surface brightness profile of the BCG + ICL up to 215 kpc in the {\em g} and {\em i} bands and found that after $\sim$75 kpc, the color profile became shallower, a clear suggestion that from this distance the ICL is starting to dominate. In addition, in concordance with previous results, the color profile indicates that the ICL formed by stripping of intermediate/massive satellites ({All the studies
involving metallicity and age quoted here have derived those quantities via stellar population synthesis models. For details, I refer the reader to the particular paper.}).

All the results reported above concern the ICL properties in intermediate/massive clusters at the present time or higher redshifts. However, numerous studies (\cite{iodice17,spavone18,raj20,ragusa21} and references therein) also focused  on the ICL in groups (also called IGL in this case) and found similar results to those mentioned above on cluster scale. \citet{spavone18} mapped the IGL of the galaxy group centered on NGC 5018, which is a local group around 40 Mpc distance, and used {\em u}, {\em g}, {\em r} images to analyze the surface brightness and color profiles of its IGL. In line with the typical behavior found for the ICL in clusters, the {\em u-g} color shows a clear negative gradient with distance, and the {\em g-r} color is consistent with that of other member galaxies. The authors concluded that the stellar stripping that is forming the IGL is still ongoing. \citet{raj20} investigated the assembly history of the BGG + IGL in the south-west group of the Fornax cluster, called Fornax A, finding no clear dependence of the BGG + IGL colors with distance but intrinsic values of them, which lead to the conclusion that the IGL formed by minor merging (likely intended as smooth accretion) and disrupted dwarf galaxies close to the group. Very recently, \citet{ragusa21} studied the properties of the IGL in the compact group of galaxies HCG 86 by deriving the surface brightness profiles (and color profiles) in the {\em g}, {\em r}, {\em i} bands, and compared their results with theoretical predictions. The fraction of the IGL and its colors, when compared with the theoretical predictions of \citet{contini14}, are consistent with the scenario where the IGL in HCG 86 formed by disruption/stripping of intermediate galaxies in the range $10<\log M_* <11$. However, when looking at the color profiles, similarly to \citet{raj20}, they did not find a clear dependence on distance.

\subsection{BCG-ICL Evolution}\label{sec:coevol}

We have seen above that the ICL is dynamically bound to the host cluster, and we\linebreak know from several studies that more massive haloes host the most massive\linebreak \mbox{BCGs (\cite{lin04,delucia07,contini12,contini14,sampaio21} and others)}, which means that both growths of ICL and BCG are linked to the host cluster. It is, then, reasonable to expect BCG and ICL to co-evolve at some time during their assembly. The key point is exactly from when the two components co-evolve. From the studies just quoted, we know that BCGs account for most of their present mass already at high redshift, while approaching the present time, more or less after $z\sim 1$, their growth rapidly
decreases. That redshift, as mentioned several times, is also the redshift at which the ICL starts to form, and its formation keeps going down to the present day. Thus, logically speaking, we should find a co-evolution of the two quantities some time after redshift $z\sim 1$.

BCGs form and evolve differently from satellite galaxies, in a hierarchical fashion via mergers and accretion, by experiencing different phases of growth: star formation at early times and a rapid growth right after due to mergers or multiple accretion of satellite galaxies (\cite{lee17} and references therein). This picture has been shown by several studies, but observationally there is still some debate on the rate at which BCGs acquire mass in the last 7--8 Gys. Some of them find no growth at all (e.g., \cite{oliva14}), others 50\% (e.g., \cite{lin13}), and others even a factor of two (e.g., \cite{lidman12}).

In \cite{contini18}, the authors extensively addressed the topic of the co-evolution of BCGs and ICL. They have shown that the fraction of the BCG stellar mass over the total within the virial radius $R_{200}$ is a decreasing function with time (see black lines in \mbox{Figure \ref{fig:contini18c}}), while the same fraction for the ICL component is, instead, an increasing function with time (see red lines in Figure \ref{fig:contini18c}). It means that the BCG stellar mass increases slower than the overall stellar mass, while the ICL, since it starts to form, acquires mass faster than the global stellar mass within the virial radius, and so these two components evolve differently and with different timescales.

\begin{figure}[H]
\includegraphics[width=13. cm]{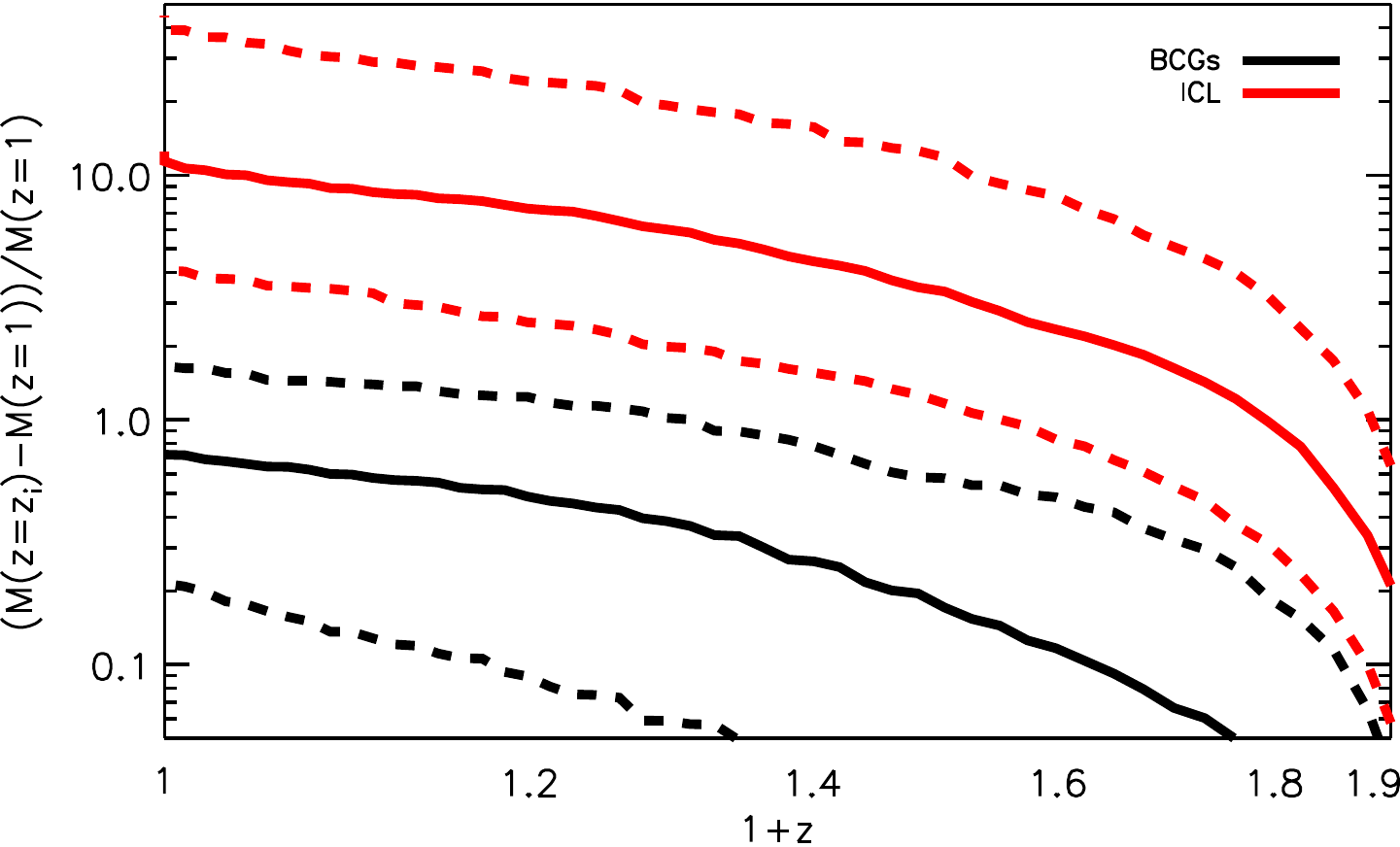}
\caption{Residual mass of the ICL (red lines) and BCGs (black lines) normalized by the mass at redshift $z=1$. Solid and dashed lines represent the median, 16th and 84th percentiles of the distributions, respectively. Credit: \citet{contini18}.}
\label{fig:contini18d}
\end{figure}

Nevertheless, it is possible that after a given redshift, BCGs and ICL grow with similar speed or similar rate. In order to verify that, we need to remove the growth that the two components do not have in common, i.e., that of the BCGs before $z\sim 1$. This is shown in Figure \ref{fig:contini18d} (from \citet{contini18}), where they show the residual in mass (the mass at any given redshift after $z=1$ and the mass at $z=1$), normalized to the mass at $z=1$ for BCGs (black lines) and ICL (red lines). The values at each redshift quantify the rate at which the two components grow, but the slopes of the two lines at each redshift quantify the speed at which the rate changes. The trend is quite clear: The slope for the ICL is very steep at the beginning of its formation but then approaches that of the BCGs, and after redshift $z\sim 0.7$, the two slopes are the same. This proves that in the last 6 or 7 Gys, the two components co-evolve.

The theoretical picture from a semi-analytic model of galaxy formation emerged above is in line with most of the observational results (e.g., \cite{burke12,burke15,morishita17,groenewald17,kluge21}). \mbox{\citet{burke15}} find that BCGs grow by a factor 1.4 in the redshift range $0.2<z<0.9$, but the factor reduces to 1.2 if they assume that 50\% of the accreted mass by the BCGs becomes ICL. Furthermore, in line with \citet{contini18}, they find that the ICL can grow by a factor of 4--5 in the same redshift range. The authors concluded that minor mergers are mainly responsible for the stellar mass assembly at late times, with the bulk of the mass  ending up in the ICL rather than the BCG (see also \citet{groenewald17}). \mbox{\citet{morishita17}} find similar results. Their ICL fraction at $z\sim 0.4$ is in the range 5--20\%, about a factor 0.5 of the present day ICL fraction, thus they concluded that BCGs and ICL have two distinct formation histories. Very recently, \citet{kluge21} studied the correlations between the absolute brightness of BCG + ICL and ICL alone with several properties linked to the host cluster, such as gravitational mass, integrated absolute brightness of satellite galaxies, their velocity dispersion, gravitational radius and others. They find that both BCG + ICL and ICL alone absolute magnitudes correlate with all of them, but in particular they correlate with the gravitational mass and satellite brightness, with the same strength. This result clearly indicates that both BCG and ICL correlate with the host cluster properties, and so both of their growths are coupled with that of the cluster. Furthermore, very recently, \mbox{\citet{furnell21}} found the ICL to grow by a factor of  2--4 in the redshift range $0.1<z<0.4$, in fairly good agreement with the growth rate found by \mbox{\citet{burke15}}, highlighting the rapid build-up of the ICL in the last few Gyr. The authors correctly concluded that the ICL is the dominant stellar evolutionary component since $z \sim 1$.

\section{ICL: An Observable Tracer of the Dark Matter Distribution}\label{sec:tracer}

The ICL, by definition, is made of stars that come from stripping (mergers) of (between) galaxies and that are not bound to any galaxy, but they feel the potential well of the cluster and so do DM particles. Moreover, DM particles are also collisionless, and it is reasonable to believe that so are stars belonging to the ICL. This similarity gave birth to the very recent idea of using the ICL to trace the DM distribution in galaxy clusters.
Although almost all the studies on the topic are just less than three years old (at the time of the writing of this review), some attempts to link the ICL to the DM are much older than that. For example, \citet{zibetti05} assumed that the ICL distribution could be approximated by using an NFW profile, while \citet{jee10} used the ICL together with weak lensing to trace DM structures in a cluster, and \citet{giallongo14} have  probably been the first to conclude that the ICL can be used to probe the DM distribution by comparing ICL observations with a DM model. Just three years later, \citet{harris17}, making use of a Fornax-like simulated cluster, found that the ICL profile is more centrally concentrated than that of the DM.

Most of the studies, however, are concentrated in  very recent years, and the first one was the work by \citet{montes19}, only because the authors used, for the first time, the title {\em Intracluster light: a luminous tracer for dark matter in clusters of galaxies}. 
With their analysis, the authors showed that the ICL follows the global DM distribution, and therefore the correct words {\em luminous tracer} were used in the title of their paper.
\citet{montes19}, with a sample of six clusters from the Hubble Frontiers Field (the same used by \citet{morishita17}), compared the distribution of the DM (obtained with strong lensing) with that of the X-ray emission of the intracluster gas, and with that of the ICL. To achieve that, they used the Modified Hausdorff distance (MHD), which is an indicator of how far two distributions are from each other. The HD is defined as the minimum distance among all the distances from a point in one set, to the closest point of the other set. The authors chose to use the modified version, which reads as follows:
\begin{equation}
d(X,Y) = \frac{1}{N_X} \sum\limits_{x_i \in X} \min\limits_{y_i \in Y} ||x_i -y_i ||  ,
\end{equation}
where $X=[x_1 ,.....x_{N_x}]$, $Y=[y_1 ,.....y_{N_y}]$, and $N_X$ is the number of elements in the set of points $X$. Compared to the classic HD, the modified version is more robust to outliers and increases monotonically with the increase in the difference between the two sets of points. {In \citet{montes19}, the sets of points
$X$ and $Y$ are the collections of any point in the ICL (or X-ray) and mass contours. As they state in their paper, an MHD value of around 20 kpc means that the average distance from the ICL contours from those of the mass is comparable with the size of the Milky Way.}
The most interesting results are essentially two: The mean MHD distance between the total matter distribution and the ICL is 25 kpc (1), and in most of the cases, the X-ray emission of the intracluster gas cannot trace the distribution of the total mass (2). Their results clearly support the idea that ICL and DM are at least connected by similar distributions, probably due to their collisionless nature, while, as we know, gas self-interacts and dissipates, and so it is highly collisional.

A hint of the link between DM and ICL came also one year earlier by the same \mbox{authors \cite{montes18}}, who found that the slope of the density profile of the BCG + ICL in their clusters was comparable to that of the total mass, remarking upon the results found by \mbox{\citet{pillepich18}} with hydrosimulations. It must be noted, however, that contemporary to \citet{montes19}, \citet{zhang19} performed a pioneering analysis on a sample of around 300 clusters at redshift $0.2<z<0.3$, and although the goals of their analysis were not as focused as those of \citet{montes19}'s analysis in the link between ICL and DM, their results suggested that the ICL is a good tracer of the cluster radial \mbox{mass distribution.}

A few months later with respect to \cite{montes19}, \citet{alonso20} repeated the same analysis performed by \citet{montes19} from a theoretical point of view, by using the Cluster-EAGLE hydrosimulations, which comprise a set of 30 zoom-in cluster simulations with remarkable spatial and gas mass resolutions. They followed the same approach, i.e., used the MHD as an indicator of how far the distributions of the total matter and ICL are, and found that the MHDs from simulations are comparable to those from observations and follow the same radial trend. Despite that they confirmed the results of \mbox{\citet{montes19}}, they also found that ICL and total matter distributions follow different radial profiles. Nevertheless, according to their findings, the ICL can be considered a good tracer of the gravitational potential.

This year (2021) has been, so far, very fruitful on this topic. \mbox{\citet{sampaio21}} explored the connection between BCG + ICL and the total cluster mass distributions by using a sample of more than 500 clusters between redshift $0.2<z<0.35$ from the Dark Energy Survey. Similarly to \citet{zhang19}, who also used a sample (but smaller) from the same survey, found that the surface brightness of the BCG + ICL has a universal radial dependence after scaling it by cluster radius. The authors also found comparable radial profiles of the BCG + ICL with those of the total cluster mass distribution (measured through weak lensing). Nevertheless, they concluded that there is not sufficient evidence to entitle the diffuse light as a good tracer of the cluster matter distribution, but it is an excellent indicator of the total cluster mass. It must be noted, however, \mbox{that \citet{montes19}} used strong lensing maps, which are more robust than the weak lensing maps used by \citet{sampaio21}. \citet{kluge21}  studied the BCG and ICL alignments with the host cluster and found that the ICL is better aligned than the BCG in terms of position angle, ellipticity and centering. This makes the ICL a good tracer of the DM distribution, including substructures. The authors also found that the ICL is more concentrated than DM, making the two radial profiles comparable. This is supported also by the model in \citet{contini20,contini21}, where the authors assumed the ICL to follow an NFW profile with a higher concentration than that of the DM. In those studies, the authors showed that such a profile for the ICL can describe the overall distribution of the ICL, since many observed scaling relations are adequately reproduced by the model.

At last, but not because of importance, it is worth mentioning another point for which the ICL can be very useful, i.e., in determining the ``splashback'' radius of a cluster. This radius is defined as the first apocenter of recently infalling DM, usually determined in simulations by looking at the radius where the slope of the density profile of the DM is the steepest (e.g., \cite{diemer14}). The splashback radius then defines a physical halo boundary. \mbox{\citet{deason21}} showed that the distribution of the ICL is directly linked to the splashback radius, which means that looking at the ICL distribution provides an independent measurement of the physical edge of a DM halo and the recent accretion rate of material. These new results have been partly confirmed by recent observations. Indeed, in their analysis of a cluster at $z\sim 0.5$ belonging to Frontiers Fields Clusters, \mbox{\citet{gonzalez21}} found a signature of the splashback radius. The authors argued that, if what they found is the splashback radius, it would be the first observed link between the ICL and the splashback radius. Clearly, more work has to be undertaken in this direction, but the idea of using the ICL to determine some physical parameters of the DM, and even its overall distribution, is surely promising.

\section{Summary and Conclusions}\label{sec:conclusions}

One of the main goals of this brief introductory review was to make clear that the ICL is an important component in galaxy groups and clusters, which can account for a relevant fraction of the whole light within the largest structures.
Not only can we not do without studying its properties because of its intrinsic importance in the context of galaxy formation and evolution, but its properties can also help to shed some light on the evolution of the structures that host the ICL. As we have seen in the last section, there are several promising applications for which the ICL can be used, and all of them need to be improved and employed in the future to increase our knowledge of the evolution of the largest structures in the Universe.

The ICL is made of stars that are not bound to any galaxy but that feels only the potential well of the host cluster. The relevant processes that contributed to its formation and evolution are essentially three: stellar stripping, galaxy mergers and pre-processing. During the last few years, there has been a growing consensus that other processes, such as disruption of dwarf satellites and {in situ} star formation, do not play an important role. Mergers and stellar stripping of intermediate mass ($10.5<\log M_* <11.5$) galaxies are the two most important contributors. From a theoretical point of view, this is expected since those kinds of galaxies are also those to experience a rapid dynamical friction and end up in the innermost regions of the host cluster faster than less massive satellites. The innermost regions are also the most concentrated where tidal forces become stronger.  Intermediate/massive galaxies as main contributors to the formation of the ICL have been predicted by theoretical studies but have also been confirmed by several observational ones. Properties such as colors, metallicity and age of the ICL, compared to the typical properties of the galaxy population, have been (and still are) very helpful. Indeed, by looking at the gradients of any of those properties, from the inner regions BCG dominated, to the outer regions ICL dominated, it has been possible to understand that the ICL is, on average, bluer, younger and less metal-rich than the BCG, with typical values closer to those of intermediate galaxies and the outermost parts of very massive galaxies.

Although there are a few examples of ICL detection in high-redshift clusters, it is quite clear that this diffuse light formed pretty late, after $z\sim 1$. It also has a different evolutionary path with respect to the BCG until a given redshift, after which the two components co-evolve. BCGs grow fast at early times and then quite slow at later times, mostly via mergers or smooth accretion. The ICL is, instead, almost non-existent before $z\sim 1$ and rapidly grows, even faster than the whole stellar matter content inside their host, at $z<1$. A co-evolution of the BCG and ICL has been shown by both theoretical and observational studies, and it is a consequence of the overall growth of the host cluster. It is shown here the importance of studying the ICL, as a tracer of the dynamical evolution of clusters.

One of the most important issues regarding the study of the ICL is surely given by the fact that it is hard to separate it from the BCG. As mentioned above, in both theoretical and observational studies, different authors use different definitions, and this makes comparisons among them difficult. However, we have made many steps forward in improving the detection of the ICL, and several techniques have been developed in order to better isolate the ICL from the rest of the light. The key point in detecting the ICL is that of removing as much contamination as possible, and several algorithms working on it are already doing a fine job. The real problem is to separate the ICL from the
BCG, which still appears a challenging task. Theoretical models can be very useful for observers in this matter. For example, \citet{contini21} investigated the typical radius at which the ICL starts to dominate, by separating a BCG + ICL system into three components: bulge, disk and ICL. The authors found that the typical radius of the transition between BCG and ICL depends on the properties of the BCG, whether or not it has an extended disk, if it is a bulge or disk like galaxy, and clearly on the amount of ICL. The radius of the transition can be as small as 15--20 kpc but can also be as large as more than 100 kpc in large BCG + ICL systems, where the disk component is important. These values are in good agreement with the first observed transition radii, such as those found in \mbox{\citet{montes21}} and \mbox{\citet{gonzalez21}}.

The link between ICL and DM has been recently addressed, and it seems to be more efficient than that with the X-ray emission of the intracluster gas. The distributions of ICL and DM within a cluster have being gradually proved to be linked together, and the technique based on the detection of ICL as a luminous tracer of the DM appears to be quite promising. More work has to be undertaken in this direction from both observational and theoretical sides to make sure that the ICL will serve as a reliable tool for studying the DM distribution in galaxies clusters. Numerical methods can provide an affordable way for making predictions that can be useful for future observational campaigns, as well as for investigating the preliminary results already obtained.

To conclude, so far, the study of the ICL has been proved to be necessary and helpful under several aspects. State-of-the-art hydrosimulations, semi-analytic models and deep observations together will surely improve our knowledge of this important component and the dynamical evolution of the structures that host it.

\vspace{6pt}


\funding{This work is supported by the National Key Research and Development Program of China (No. 2017YFA0402703) and by the National Natural Science Foundation of China (Key Project No. 11733002).}

\dataavailability{Data used in this review are available under request to the corresponding
author of each study.} 


\conflictsofinterest{The author declares no conflict of interest.}


\end{paracol}
\reftitle{References}



\begin{thebibliography}{999}
\bibitem[Zwicky F. (1937)]{zwicky37} Zwicky, F. On the Masses of Nebulae and of Clusters of Nebulae. {\em \apj} {\bf 1937}, {\em 86}, 217. [\href{http://doi.org/10.1086/143864}{CrossRef}]
\bibitem[Montes M. (2019)]{montes19rev} Montes, M. The Intracluster Light and its Role in Galaxy Evolution in Clusters. {\em arXiv} {\bf 2019}, {arXiv:1912.01616}.
\bibitem[Gonzalez et al.( 2013)]{gonzalez13} Gonzalez, A.H.; Sivanandam, S.; Zabludoff, A.I.; Zaritsky, D.
Galaxy Cluster Baryon Fractions Revisited. {\em \apj} {\bf 2013}, {\em 778}, 14. [\href{http://dx.doi.org/10.1088/0004-637X/778/1/14}{CrossRef}]
\bibitem[Contini et al. (2014)]{contini14} Contini, E.; De Lucia, G.; Villalobos, {\'A}.; Borgani, S. On the Formation and Physical Properties of the Intracluster Light in Hierarchical Galaxy Formation Models. {\em \mnras} {\bf 2014}, {\em 437}, 3787--3802. [\href{http://dx.doi.org/10.1093/mnras/stt2174}{CrossRef}]
\bibitem[Presotto et al. (2014)]{presotto14} Presotto, V.; Girardi, M.; Nonino, M.;  Mercurio, A.; Grillo, C.; Rosati, P.; Biviano, A.; Annunziatella, M.; Balestra, I.; Cui, W.; et al. Intracluster Light Properties in the CLASH-VLT Clusters MACS J1206.2-0847. {\em \aap} {\bf 2014}, {\em 565}, A126. [\href{http://dx.doi.org/10.1051/0004-6361/201323251}{CrossRef}]
\bibitem[Contini et al. (2018)]{contini18} Contini, E.; Yi, S.K.; Kang, X. The Different Growth Pathways of Brightest Cluster Galaxies and Intracluster Light. {\em \mnras} {\bf 2018}, {\em 479}, 932--944.
\bibitem[Arnaboldi et al. (2012)]{arnaboldi12} Arnaboldi, M.; Ventimiglia, G.; Iodice, E.; Gerhard, O.  A Tale of two Tails and an Off-Centered Envelope: Diffuse Light Around the cD Galaxy NGC3311 in the Hydra I Cluster. {\em \aap} {\bf 2012}, {\em 545}, A37. [\href{http://dx.doi.org/10.1051/0004-6361/201116752}{CrossRef}]
\bibitem[Edwards et al. (2016)]{edwards16} Edwards, L.O.V.; Alpert, H.S.; Trierweiler, I.L.;  Abraham, T.; Beizer, V.G.    Stellar Populations of BCGs, Close Companions and Intracluster Light in Abell85, Abell2457 and IIZw108. {\em \mnras} {\bf 2016}, {\em 461}, 230--239. [\href{http://dx.doi.org/10.1093/mnras/stw1314}{CrossRef}]
\bibitem[Kravtsov et al. (2018)]{kravtsov18} Kravtsov, A.V.; Vikhlinin, A.A.; Meshcheryakov, A.V. Stellar Mass-Halo Mass Relation and Star Formation Efficiency in High-Mass Halos. {\em Astron. Lett.} {\bf 2018}, {\em 44}, 8--34. [\href{http://dx.doi.org/10.1134/S1063773717120015}{CrossRef}]
\bibitem[Zhang et al. (2019)]{zhang19} Zhang, Y.; Yanny, B.; Palmese, A.; Gruen, D.; To, C.; Rykoff, E.S.; Leung, Y.; Collins, C.; Hilton, M.; Abbott, T.M.C.; et al.  Dark Energy Survey Year 1 Results: Detection of Intracluster Light at Redshift $\sim 0.25$. {\em \apj} {\bf 2019}, {\em 874}, 165. [\href{http://dx.doi.org/10.3847/1538-4357/ab0dfd}{CrossRef}]
\bibitem[DeMaio et al.(2020)]{demaio20} DeMaio, T.; Gonzalez, A.H.; Zabludoff, A.; Zaritsky, D.; Aldering, G.; Brodwin, M; Connor, T.; Donahue, M.; Hayden, B.; Mulchaey, J.S.; et al. The Growth of Brightest Cluster Galaxies and Intracluster Light Over the Past 10 Billion Years. {\em \mnras} {\bf 2020}, {\em 491}, 3751--3759. [\href{http://dx.doi.org/10.1093/mnras/stz3236}{CrossRef}]
\bibitem[Montes et al. (2021)]{montes21} Montes, M.; Brough, S.; Owers, M.; Santucci, G.  The Buildup of the Intracluster Light of A85 as Sees by Subaru's Hyper Suprime-Cam. {\em \apj} {\bf 2021}, {\em 910}, 45. [\href{http://dx.doi.org/10.3847/1538-4357/abddb6}{CrossRef}]
\bibitem[Rudick et al. (2011)]{rudick11} Rudick, C.S.; Mihos, J.C.; McBride, C.K.    The Quantity of Intracluster Light: Comparing Theoretical and Observational Measurement Techiques Using Simulated Clusters. {\em \apj} {\bf 2011}, {\em 732}, 48. [\href{http://dx.doi.org/10.1088/0004-637X/732/1/48}{CrossRef}]
\bibitem[Cui et al. (2014)]{cui14} Cui, W.; Murante, G.; Monaco, P.; Borgani, S.; Granato, G.L.; Killedar, M.; Lucia, G.D.; Presotto, V.; Dolag, K. Characterizing Diffused Stellar Light in Simulated Galaxy Clusters. {\em \mnras} {\bf 2014}, {\em 437}, 816--830. [\href{http://dx.doi.org/10.1093/mnras/stt1940}{CrossRef}]
\bibitem[Jimenez-Teja; Dupke (2016)]{jimenez16} Jimenez-Teja,  Y.; Dupke, R. Disentangling the ICL with the CHEFs: Abell 2744 as a Case Study. {\em \apj} {\bf 2016}, {\em 820}, 49. [\href{http://dx.doi.org/10.3847/0004-637X/820/1/49}{CrossRef}]
\bibitem[Montes; Trujillo (2018)]{montes18} Montes, M.; Trujillo, I. Intracluster Light at the Frontier - II. The Frontier Fields Clusters. {\em \mnras} {\bf 2018}, {\em 474}, 917--932. [\href{http://dx.doi.org/10.1093/mnras/stx2847}{CrossRef}]
\bibitem[Tang et al. (2018)]{tang18} Tang, L.; Lin, W.; Cui, W.; Kang, X.; Wang, Y.; Contini, E.; Yu, Y.  An Investigation of Intracluster Light Evolution Using Cosmological Hydrodynamical Simulations. {\em \apj} {\bf 2018}, {\em 859}, 85. [\href{http://dx.doi.org/10.3847/1538-4357/aabd78}{CrossRef}]
\bibitem[Dolag et al. (2010)]{dolag10} Dolag, K.; Murante, G.; Borgani, S. Dynamical Difference Between the cD Galaxy and the Diffuse, Stellar Component in Simulated Galaxy Clusters. {\em \mnras} {\bf 2010}, {\em 405}, 1544--1559. [\href{http://dx.doi.org/10.1111/j.1365-2966.2010.16583.x}{CrossRef}]
\bibitem[Somerville et al. (2008)]{somerville08} Somerville, R.S.; Hopkins, P.E.; Cox, T.J.; Robertson, B.E.; Hernquist, L. A Semi-Analytic Model for the Co-Evolution of Galaxies, Black Holes and Active Galactic Nuclei. {\em \mnras} {\bf 2008}, {\em 391}, 481--506. [\href{http://dx.doi.org/10.1111/j.1365-2966.2008.13805.x}{CrossRef}]
\bibitem[Henriques; Thomas (2010)]{henriques10} Henriques, B.M.B.; Thomas, P.A. Tidal Disruption of Satellite Galaxies in a Semi-Analytic Model of Galaxy Formation. {\em \mnras} {\bf 2010}, {\em 403}, 768--779. [\href{http://dx.doi.org/10.1111/j.1365-2966.2009.16151.x}{CrossRef}]
\bibitem[Guo et al. (2011)]{guo11} Guo, Q.; White, S.; Boylan-Kolchin, M.; Lucia, G.;  Kauffmann, G.; Lemson, G.; Li, C.; Springel, V.; Weinmann, S. From Dwarf Spheroidals to cD Galaxies: Simulating the Galaxy Population in a $\Lambda$CDM Cosmology. {\em \mnras} {\bf 2011}, {\em 413}, 101--131.
\bibitem[Feldmeier et al. (2004)]{feldmeier04} Feldmeier, J.; Mihos, J.C.; Morrison, H.; Harding, P.; Kaib, N.; Dubinski, J. Deep CCD Surface Photometry of Galaxy Clusters II. Searching for Intracluster Starlight in Non-cD clusters. {\em \apj} {\bf 2004}, {\em 609}, 617--637. [\href{http://dx.doi.org/10.1086/421313}{CrossRef}]
\bibitem[Zibetti et al. (2005)]{zibetti05} Zibetti, S.; White, S.; Schneider, D.P.; Brinkmann, J. Intergalactic Stars in $z\sim 0.25$ Galaxy Clusters: Systematic Properties from Stacking of Sloan Digital Sky Survey Imaging Data. {\em \mnras} {\bf 2005}, {\em 358}, 949--967. [\href{http://dx.doi.org/10.1111/j.1365-2966.2005.08817.x}{CrossRef}]
\bibitem[Furnell et al. (2021)]{furnell21} Furnell, K.E.; Collins, C.A.; Kelvin, L.S.; Baldry, I.K.; James, P.J.; Manolopoulou, M.; Mann, R.G.; Giles, P.A.; Bermeo, A.; Hilton, M.  The Growth of Intracluster Light in XCS-HSC Galaxy Clusters from $0.1<z<0.5$. {\em \mnras} {\bf 2021}, {\em 502}, 2419--2437.
\bibitem[Sersic (1968)]{sersic68} Sersic, J.L. \emph{Cordoba, Argentina: Observatorio Astronomico};
Universidad Nacional de Córdoba: Córdoba, Argentina, 1968.
\bibitem[Gonzalez et al. (2005)]{gonzalez05} Gonzalez, A.; Zabludoff, A.I.; Zaritsky, D. Intracluster Light in Nearby Galaxy Cluster: Relationship to the Halos of Brightest Cluster Galaxies. {\em \apj} {\bf 2005}, {\em 618}, 195--213. [\href{http://dx.doi.org/10.1086/425896}{CrossRef}]
\bibitem[Seigar et al. (2007)]{seigar07} Seigar, M.S.; Graham, A.W.; Jerjen, H. Intracluster Light and the Extended Stellar Envelopes of cD Galaxies: An Analytical Description. {\em \mnras} {\bf 2007}, {\em 378}, 1575--1588. [\href{http://dx.doi.org/10.1111/j.1365-2966.2007.11899.x}{CrossRef}]
\bibitem[Donzelli et al. (2011)]{donzelli11} Donzelli, C.J.; Muriel, H.; Madrid, J.P. The Luminosity Profiles of Brightest Cluster Galaxies. {\em \apjs} {\bf 2011}, {\em 195}, 15. [\href{http://dx.doi.org/10.1088/0067-0049/195/2/15}{CrossRef}]
\bibitem[Giallongo et al. (2014)]{giallongo14} Giallongo, E.; Menci, N.; Grazian, A.; Gallozzi. S.; Castellano, M.; Fiore, F.; Fontana, A.; Pentericci, L.; Boutsia, K.; Paris, D. Diffuse Optical Intracluster Light as a Measure of Stellar Tidal Stripping: The Cluster CL0024+17 at $z\sim 0.4$ Observed at the Large Binocular Telescope. {\em \apj} {\bf 2015}, {\em 781}, 24. [\href{http://dx.doi.org/10.1088/0004-637X/781/1/24}{CrossRef}]
\bibitem[Cooper et al. (2015)]{cooper15} Cooper, A.P.; Gao, L.; Guo, Q.; Frenk, C.S.; Jenkins, A.   Surface Photometry of Brightest Cluster Galaxies and Intracluster Stars in $\Lambda$CDM. {\em \mnras} {\bf 2015}, {\em 451}, 2703--2722.
\bibitem[Alamo-Martinez; Blakeslee (2017)]{alamo17} Alamo-Martinez, K.A.; Blakeslee, J.P. Specific Frequencies and Luminosity Profiles of Cluster Galaxies and Intracluster Light in Abell1689. {\em \apj} {\bf 2017}, {\em 849}, 6. [\href{http://dx.doi.org/10.3847/1538-4357/aa8f44}{CrossRef}]
\bibitem[Durret et al. (2019)]{durret19} Durret, F.; Tarricq, Y.; Marquez, I.; Ashkar, H.; Adami, C. Link Between Brightest Cluster Galaxy Properties and Large Scale Extensions of 38 DAFT/FADA and CLASH Clusters in the Redshift Range $0.2<z<0.9$. {\em \aap} {\bf 2019}, {\em 622}, A78.
\bibitem[Kluge et al. (2021)]{kluge21} Kluge, M.; Bender, R.; Riffeser, A.; Goessl, C.; Hopp, U.; Schmidt, M.; Ries, C. Photometric Dissection of Intracluster Light and Its Correlations with Host Cluster Properties. {\em \apjs} {\bf 2021}, {\em 252}, 27. [\href{http://dx.doi.org/10.3847/1538-4365/abcda6}{CrossRef}]
\bibitem[Schuberth et al.(2008)]{schuberth08} Schuberth, Y.; Richtler, T.; Bassino, L.; Hilker, M.  Intra-cluster globular clusters around NGC 1399 in Fornax? {\em \aap} {\bf 2008}, {\em 477}, L9. [\href{http://dx.doi.org/10.1051/0004-6361:20078668}{CrossRef}]
\bibitem[D'Abrusco et al. (2016)]{dabrusco16} D'Abrusco, R.; Cantiello, M.; Paolillo, M.; Pota, V.; Napolitano, N.R.; Limatola, L.; Spavone, M.; Grado, A.; Iodice, E.; Capaccioli, M.  The Extended Spatial Distribution of Globular Clusters in the Core of the Fornax Cluster.
{\em  Astrophys. J. Lett.} {\bf 2016}, {\em 819}, L31. [\href{http://dx.doi.org/10.3847/2041-8205/819/2/L31}{CrossRef}]
\bibitem[Cantiello et al.(2020)]{cantiello20} Cantiello, M.; Venhola, A.; Grado, A.; Paolillo, M.; D’Abrusco, R; Raimondo, G.; Quintini, M.; Hilker, K.; Mieske, S.; Tortora, C.; et al. The Fornax Deep Survey with VST. IX. Catalog of sources in the FDS area with an example study for globular clusters and background galaxies. {\em \aap} {\bf 2020}, {\em 639}, A136. [\href{http://dx.doi.org/10.1051/0004-6361/202038137}{CrossRef}]
\bibitem[Lee et al.(2010)]{lee10} Lee, M.G.; Park, H.S.; Hwang, H.S. Detection of a Large-Scale Structure of Intracluster Globular Clusters in the Virgo Cluster. {\em Science} {\bf 2010}, {\em 328}, 334. [\href{http://dx.doi.org/10.1126/science.1186496}{CrossRef}] [\href{http://www.ncbi.nlm.nih.gov/pubmed/20223950}{PubMed}]
\bibitem[Durrell et al.(2014)]{durrell14} Durrell, P.R.; C{\^o}t{\'e}, P.; Peng, E.W.; Blakeslee, J. The Next Generation Virgo Cluster Survey. VIII. The Spatial Distribution of Globular Clusters in the Virgo Cluster. {\em \apj} {\bf 2014}, {\em 794}, 103. [\href{http://dx.doi.org/10.1088/0004-637X/794/2/103}{CrossRef}]
\bibitem[Ko et al.(2017)]{ko17} Ko, Y.; Hwang, H.S.; Lee, M.G.; Park, H.S.; Lim, S.; Sohn, J.; Jang, I.S.; Hwang, N.; Park B.G. To the Edge of M87 and Beyond: Spectroscopy of Intracluster Globular Clusters and Ultracompact Dwarfs in the Virgo Cluster. {\em \apj} {\bf 2017}, {\em 835}, 212. [\href{http://dx.doi.org/10.3847/1538-4357/835/2/212}{CrossRef}]
\bibitem[Longobardi et al.(2018)]{longobardi18} Longobardi, A.; Peng, E.W.; C{\^o}t{\'e}, P.; Mihos, J.C.; Ferrarese, L.; Puzia, T. H.; Lançon, A.; Zhang, H.; Muñoz, R.P.; Blakeslee, J.P.;  et~al. The Next Generation Virgo Cluster Survey (NGVS). XXXI. The Kinematics of Intracluster Globular Clusters in the Core of the Virgo Cluster. {\em \apj} {\bf 2018}, {\em 864}, 36. [\href{http://dx.doi.org/10.3847/1538-4357/aad3d2}{CrossRef}]
\bibitem[Harris et al.(2020)]{harris20} Harris, W.E.; Brown, R.A.; Durrell, P.R.; Romanowsky, A.J.; Blakeslee, J.; Brodie, J.; Janssens, S.; Lisker, T.; Okamoto, S.; Wittmann, S.; et al. The PIPER Survey. I. An Initial Look at the Intergalactic Globular Cluster Population in the Perseus Cluster. {\em \apj} {\bf 2020}, {\em 890}, 105. [\href{http://dx.doi.org/10.3847/1538-4357/ab6992}{CrossRef}]
\bibitem[Peng et al.(2011)]{peng11} Peng, E.W.; Ferguson, H.C.; Goudfrooij, P.;  Hammer, D.; Lucey, J.R.; Marzke, R.O.; Puzia, T.H.; Carter, D.; Balcells, M.; Bridges, T.; et al. The HST/ACS Coma Cluster Survey. IV. Intergalactic Globular Clusters and the Massive Globular Cluster System at the Core of the Coma Galaxy Cluster. {\em \apj} {\bf 2011}, {\em 730}, 23. [\href{http://dx.doi.org/10.1088/0004-637X/730/1/23}{CrossRef}]
\bibitem[Madrid et al.(2018)]{madrid18} Madrid, J.P.; O'Neill, C.R.; Gagliano, A.T.; Marvil, J.R.  A Wide-field Map of Intracluster Globular Clusters in Coma. {\em \apj} {\bf 2018}, {\em 867}, 144. [\href{http://dx.doi.org/10.3847/1538-4357/aae206}{CrossRef}]
\bibitem[West et al.(2011)]{west11} West, M.J.; Jord{\'a}n, A.; Blakeslee, J.P.; Côté, P.; Gregg, M.D.; Takamiya, M.; Marzke, R.O.  The globular cluster systems of Abell 1185. {\em \aap} {\bf 2011}, {\em 528}, A115. [\href{http://dx.doi.org/10.1051/0004-6361/201015939}{CrossRef}]
\bibitem[Longobardi et al.(2015)]{longobardi15} Longobardi, A.; Arnaboldi, M.; Gerhard, O.; Hanuschik, R.   The outer regions of the giant Virgo galaxy M 87 Kinematic separation of stellar halo and intracluster light. {\em \aap} {\bf 2015}, {\em 579}, A135. [\href{http://dx.doi.org/10.1051/0004-6361/201525773}{CrossRef}]
\bibitem[Kelson et al.(2002)]{kelson02} Kelson, D.D.; Zabludoff, A.I.; Williams, K.A.; Trager, S.C.; Mulchaey, J.S.; Bolte, M.  Determination of the Dark Matter Profile of A2199 from Integrated Starlight. {\em \apj} {\bf 2002}, {\em 576}, 720. [\href{http://dx.doi.org/10.1086/341891}{CrossRef}]
\bibitem[Bender et al.(2015)]{bender15} Bender, R.; Kormendy, J.; Cornell, M.E.;  Fisher, D.B.  Structure and Formation of cD Galaxies: NGC 6166 in ABELL 2199. {\em \apj} {\bf 2015}, {\em 807}, 56. [\href{http://dx.doi.org/10.1088/0004-637X/807/1/56}{CrossRef}]
\bibitem[Monaco et al. (2006)]{monaco06} Monaco, P.; Murante, G.; Borgani, S.; Fontanot, F. Diffuse Stellar Component in Galaxy Clusters and the Evolution of the Most Massive Galaxies at $z<1$. {\em \apj} {\bf 2006}, {\em 652} L89--L92.
\bibitem[Murante et al. (2007)]{murante07} Murante, G.; Giovalli, M.; Gerhard, O.; Arnaboldi, M.; Borgani , S.; Dolag, K. The Importance of Mergers for the Origin of Intracluster Stars in Cosmological Simulations of Galaxy Clusters. {\em \mnras} {\bf 2007}, {\em 377}, 2--16. [\href{http://dx.doi.org/10.1111/j.1365-2966.2007.11568.x}{CrossRef}]
\bibitem[Gerhard et al., (2007)]{gerhard07} Gerhard, O.; Arnaboldi, M.; Freeman, K.C.; Okamura, S.; Kashikawa, N.; Yasuda, N.  The Kinematics of Intracluster Planetary Nebulae and the On-Going Subcluster Merger in the Coma Cluster Core. {\em \aap} {\bf 2007}, {\em 468}, 815--822. [\href{http://dx.doi.org/10.1051/0004-6361:20066484}{CrossRef}]
\bibitem[Burke et al. (2015)]{burke15} Burke, C.; Hilton, M.; Collins, C. Coevolution of Brightest Cluster Galaxies and Intracluster Light Using CLASH. {\em \mnras} {\bf 2015}, {\em 449}, 2353--2367. [\href{http://dx.doi.org/10.1093/mnras/stv450}{CrossRef}]
\bibitem[Groenewald et al. (2017)]{groenewald17} Groenewald, D.N.; Skelton, R.E.; Gilbank, D.G.; Loubser, S.I. The Close Pair Fraction of BCGs Since $z=0.5$: Major Mergers Dominate Recent BCG Stellar Mass Growth. {\em \mnras} {\bf 2017}, {\em 467}, 4101--4117. [\href{http://dx.doi.org/10.1093/mnras/stx340}{CrossRef}]
\bibitem[Jimenez-Teja et al. (2018)]{jimenez18} Jimenez-Teja, Y.; Dupke, R.; Benitez, N.; Koekemoer, A.M.; Zitrin, A.; Umetsu, K.;  Ziegler, B.L.; Frye, B.L.; Ford, H.; Bouwens, R.J.; et al. Unveiling the Dynamical State of Massive Clusters Through the ICL Fraction. {\em \apj} {\bf 2018}, {\em 857}, 19. [\href{http://dx.doi.org/10.3847/1538-4357/aab70f}{CrossRef}]
\bibitem[Jimenez-Teja et al. (2019)]{jimenez19} Jimenez-Teja, Y.; Dupke, R.; Lopes de Oliveira, R.; Xavier, H.S.; Coelho, P.R.T.; Chies-Santos, A.L.; López-Sanjuan, C.; Alvarez-Candal, A.; Costa-Duarte, M.V.; Telles, E.; et al. J-PLUS: Analysis of the Intracluster Light in the Coma Cluster. {\em \aap} {\bf 2019}, {\em 622}, A183. [\href{http://dx.doi.org/10.1051/0004-6361/201833547}{CrossRef}]
\bibitem[Purcell et al.(2007)]{purcell07} Purcell, C.W.; Bullock, J.S.; Zentner, A.R. Shredded Galaxies as the Source of Diffuse Intrahalo Light on Varying Scales. {\em \apj} {\bf 2007}, {\em 666}, 20--33. [\href{http://dx.doi.org/10.1086/519787}{CrossRef}]
\bibitem[Conroy et al. (2007)]{conroy07} Conroy, C.; Wechsler, R.H.; Kravtsov, A.V. The Hierarchical Build-Up of Massive Galaxies and the Intracluster Light Since $z=1$. {\em \apj} {\bf 2007}, {\em 668}, 826--838. [\href{http://dx.doi.org/10.1086/521425}{CrossRef}]
\bibitem[Martel et al. (2012)]{martel12} Martel, H.; Barai, P.; Brito, W. The Fate of Dwarf Galaxies in Cluster and the Origin of Intracluster Stars II. Cosmological Simulations. {\em \apj} {\bf 2012}, {\em 757}, 48. [\href{http://dx.doi.org/10.1088/0004-637X/757/1/48}{CrossRef}]
\bibitem[Annunziatella et al. (2016)]{annunziatella16} Annunziatella, M.; Mercurio, A.; Biviano, A.; Girardi, M.; Nonino, M.; Balestra, I.; Rosati, P.; Caminha, G.B.; Brescia, M.; Gobat, R.; et al. CLASH-VLT: Environmental-driven Evolution of Galaxies in the $z=0.209$ Cluster Abell 209. {\em \aap} {\bf 2016}, {\em 585}, A160. [\href{http://dx.doi.org/10.1051/0004-6361/201527399}{CrossRef}]
\bibitem[Morishita et al.( 2017)]{morishita17} Morishita, T.; Abramson, L.E.; Treu, T.; Schmidt, K.B. ; Vulcani, B.; Wang, X. Characterizing Intracluster Light in the Hubble Frontier Fields. {\em \apj} {\bf 2017}, {\em 846}, 139. [\href{http://dx.doi.org/10.3847/1538-4357/aa8403}{CrossRef}]
\bibitem[Raj et al. (2020)]{raj20} Raj, M.A.; Iodice, E.; Napolitano, N.R.; Hilker, M.; Spavone, M.; Peletier, R.F.; Su, H-S.; Falcón-Barroso, J.; Ven, G.v.d.; Cantiello, M.; et al. The Fornax Deep Survey with VST. X. The Assembly History of the Bright Galaxies and Intra-Group Light in the Fornax A Subgroup. {\em \aap} {\bf 2020}, {\em 640}, A137. [\href{http://dx.doi.org/10.1051/0004-6361/202038043}{CrossRef}]
\bibitem[Rudick et al. (2009)]{rudick09} Rudick, C.S.; Mihos, J.C.; Frey, L.H.; McBride, C.K. Tidal Streams of Intracluster Light. {\em \apj} {\bf 2009}, {\em 699}, 1518--1529. [\href{http://dx.doi.org/10.1088/0004-637X/699/2/1518}{CrossRef}]
\bibitem[DeMaio et al. (2015)]{demaio15} DeMaio, T.; Gonzalez, A.H.; Zabludoff, A.; Zaritsky, D.B. On the Origin of the Intracluster Light in Massive Galaxy Clusters. {\em \mnras} {\bf 2015}, {\em 448}, 1162--1177. [\href{http://dx.doi.org/10.1093/mnras/stv033}{CrossRef}]
\bibitem[DeMaio et al. (2018)]{demaio18} DeMaio, T.; Gonzalez, A.H.; Zabludoff, A.; Zaritsky, D.; Connor, T.; Donahue, M.; Mulchaey, J.S. Lost but not Forgotten: Intracluster Light in Galaxy Groups and Clusters. {\em \mnras} {\bf 2018}, {\em 474}, 3009--3031. [\href{http://dx.doi.org/10.1093/mnras/stx2946}{CrossRef}]
\bibitem[Willman et al. (2004)]{willman04} Willman, B.; Governato, F.; Wadsley, J.; Quinn, T. The Origin and Properties of Intracluster Stars in a Rich Cluster. {\em \mnras} {\bf 2004}, {\em 355}, 159--168. [\href{http://dx.doi.org/10.1111/j.1365-2966.2004.08312.x}{CrossRef}]
\bibitem[Mihos et al. (2005)]{mihos05} Mihos, J.C.; Harding, P.; Feldmeier, J.; Morrison, H. Diffuse Light in the Virgo Cluster. {\em \apj} {\bf 2005}, {\em 631}, L41--L44. [\href{http://dx.doi.org/10.1086/497030}{CrossRef}]
\bibitem[Rudick et al. (2006)]{rudick06} Rudick, C.S.; Mihos, J.C.; McBride, C. The Formation and Evolution of Intracluster Light. {\em \apj} {\bf 2006}, {\em 648}, 936--946. [\href{http://dx.doi.org/10.1086/506176}{CrossRef}]
\bibitem[Sommer-Larsen J.(2006)]{sommer-larsen06} Sommer-Larsen, J. Properties of Intra-group Stars and Galaxies in Galaxy Groups: ``Normal'' versus ``Fossil'' Groups. {\em \mnras} {\bf 2006}, {\em 369}, 958--968. [\href{http://dx.doi.org/10.1111/j.1365-2966.2006.10352.x}{CrossRef}]
\bibitem[Puchwein et al. (2010)]{puchwein10} Puchwein, E.; Springel, V.; Sijacki, D.; Dolag, K. Intracluster Stars in Simulations with Active Galactic Nucleus Feedback. {\em \mnras} {\bf 2010}, {\em 406}, 936--951. [\href{http://dx.doi.org/10.1111/j.1365-2966.2010.16786.x}{CrossRef}]
\bibitem[Melnick et al. (2012)]{melnick12} Melnick, J.; Giraud, E.; Toledo, I.;  Selman, F.;  Quintana, H. Intergalactic Stellar Populations in Intermediate Redshift Clusters. {\em \mnras} {\bf 2012}, {\em 427}, 850--858. [\href{http://dx.doi.org/10.1111/j.1365-2966.2012.21924.x}{CrossRef}]
\bibitem[Montes et al. (2014)]{montes14} Montes, M.; Trujillo, I. Intracluster Light at the Frontier: A2744. {\em \apj} {\bf 2014}, {\em 794}, 137. [\href{http://dx.doi.org/10.1088/0004-637X/794/2/137}{CrossRef}]
\bibitem[Krick et al. (2006)]{krick06} Krick, J.E.; Bernstein, R.A.; Pimbblet, K.A. Diffuse Optical Light in Galaxy Clusters. I. Abell 3888. {\em \apj} {\bf 2006}, {\em 131}, 168--184. [\href{http://dx.doi.org/10.1086/498269}{CrossRef}]
\bibitem[Krick et al. (2007)]{krick07} Krick, J.E.; Bernstein, R.A. Diffuse Optical Light in Galaxy Clusters. II. Correlations with Cluster Properties. {\em \apj} {\bf 2007}, {\em 134}, 466--493. [\href{http://dx.doi.org/10.1086/518787}{CrossRef}]
\bibitem[Rudick et al. (2010)]{rudick10} Rudick, C.S.; Mihos, J.C.; Harding, P.;  Feldmeier, J.J.; Janowiecki, S.; Morrison, H.L. Optical Colors of Intracluster Light in the Virgo Cluster. {\em \apj} {\bf 2010}, {\em 720}, 569--580. [\href{http://dx.doi.org/10.1088/0004-637X/720/1/569}{CrossRef}]
\bibitem[Toledo et al. (2011)]{toledo11} Toledo, I.; Melnick, J.; Selman, F.; Quintana, H.; Giraud, E.; Zelaya, P. Diffuse Intracluster Light at Intermediate Redshifts: Intracluster Light Observations in an X-ray Cluster at $z=0.29$. {\em \mnras} {\bf 2011}, {\em 414} 602--614. [\href{http://dx.doi.org/10.1111/j.1365-2966.2011.18423.x}{CrossRef}]
\bibitem[Iodice et al. (2017)]{iodice17} Iodice, E.; Spavone, M.; Cantiello, M.; DAbrusco, R.; Capaccioli, M.; Hilker, M.; Mieske, S.; Napolitano, N.; Peletier, R.;  Limatola, L.; et al. Intracluster Patches of Baryons in the Core of the Fornax Cluster. {\em \apj} {\bf 2017}, {\em 851}, 75. [\href{http://dx.doi.org/10.3847/1538-4357/aa9b30}{CrossRef}]
\bibitem[Harris et al. (2017)]{harris17} Harris, K.A.; Debattista, V.P.; Governato, F.; Thompson, B.B.; Clarke, A.J.; Quinn, T.; Willman, B.; Benson, A.; Farrah, D.; Peng, E.W.; et al. Quantifying the Origin and Distribution of Intracluster Light in a Fornax-Like Cluster. {\em \mnras} {\bf 2017}, {\em 467}, 4501--4513. [\href{http://dx.doi.org/10.1093/mnras/stx401}{CrossRef}]
\bibitem[Contini et al. (2019)]{contini19} Contini, E.; Yi, S.K.; Kang, X. Theoretical Predictions of Colors and Metallicity of the Intracluster Light. {\em \apj} {\bf 2019}, {\em 871}, 24. [\href{http://dx.doi.org/10.3847/1538-4357/aaf41f}{CrossRef}]
\bibitem[Spavone et al. (2020)]{spavone20} Spavone, M.; Iodice, E.; van de Ven, G.; Falcón-Barroso,  J.; Raj, M.A.; Hilker, M.; Peletier, R.P.; Capaccioli, M.; Mieske, S.; Venhola, A.; et al. The Fornax Deep Survey with VST. VIII. Connecting the Accretion History with the Cluster Density. {\em \aap} {\bf 2020}, {\em 639}, A14. [\href{http://dx.doi.org/10.1051/0004-6361/202038015}{CrossRef}]
\bibitem[Edwards et al. (2020)]{edwards20} Edwards, L.O.V.; Salinas, M.; Stanley, S.; West, P.E.H.; Trierweiler, I.;	Alpert, H.; Coelho, P.; Koppaka, S.; Tremblay, G.R.; Martel, H.; et al. Clocking the Formation of Today's Largest Galaxies: Wide Field Spectroscopy of Brightest Cluster Galaxies and Their Surroundings. {\em \mnras} {\bf 2020}, {\em 491}, 2617--2638. [\href{http://dx.doi.org/10.1093/mnras/stz2706}{CrossRef}]
\bibitem[Gu et al. (2020)]{gu20} Gu, M.; Conroy, C.; Law, D.;  Dokkum, P.V.; Yan, R.; Wake, D.; Bundy, K.; Villaume, A.; Abraham, R.; Merritt, A.; et al.  Spectroscopic Constraints on the Buildup of Intracluster Light in the Coma Cluster. {\em \apj} {\bf 2020}, {\em 894}, 32. [\href{http://dx.doi.org/10.3847/1538-4357/ab845c}{CrossRef}]
\bibitem[Doherty et al. (2009)]{doherty09} Doherty, M.; Arnaboldi, M.; Das, P.; Gerhard, O.; Aguerri, J.A.L.; Ciardullo, R.; Feldmeier, J.J.; Freeman, K.C.; Jacoby, G.H.; Murante, G.; et al. The Edge of the M87 Halo and the Kinematics of the Diffuse Light in the Virgo Cluster Core. {\em \aap} {\bf 2009}, {\em 502}, 771--786. [\href{http://dx.doi.org/10.1051/0004-6361/200811532}{CrossRef}]
\bibitem[Ventimiglia et al. (2011)]{ventimiglia11} Ventimiglia, G.; Arnaboldi, M.; Gerhard, O. The Unmixed Kinematics and Origins of the Diffuse Stellar Light in the Core of the Hydra I Cluster (Abell1060). {\em \aap} {\bf 2011}, {\em 528}, A24. [\href{http://dx.doi.org/10.1051/0004-6361/201015982}{CrossRef}]
\bibitem[Pillepich et al. (2018)]{pillepich18} Pillepich, A.; Nelson, D.; Hernquist, L.; Springel, V.; Pakmor, R.; Torrey, P.; Weinberger, R.; Genel, S.; Naiman, J.P.; Marinacci, F.; et al. First Results from the IllustrisTNG Simulations: The Stellar Mass Content of Groups and Clusters of Galaxies. {\em \mnras} {\bf 2018}, {\em 475}, 648--675. [\href{http://dx.doi.org/10.1093/mnras/stx3112}{CrossRef}]
\bibitem[Montes; Trujillo (2019)]{montes19} Montes, M.; Trujillo, I. Intracluster Light: A Luminous Tracer for Dark Matter in Cluster of Galaxies. {\em \mnras} {\bf 2019}, {\em 482}, 2838--2851. [\href{http://dx.doi.org/10.1093/mnras/sty2858}{CrossRef}]
\bibitem[Alonso Asensio et al. (2020)]{alonso20} Alonso Asensio, I.; Dalla Vecchia, C.; Bah{\'e}, Y.M.;  Barnes, D.J.; Kay, S.T.  The Intracluster Light as a Tracer of the Total Matter Density Distribution: A View from Simulations. {\em \mnras} {\bf 2020}, {\em 494}, 1859--1864. [\href{http://dx.doi.org/10.1093/mnras/staa861}{CrossRef}]
\bibitem[Kluge et al. (2020)]{kluge20} Kluge, M.; Neureiter, B.; Riffeser, A.; Bender, R.; Goessl, C.; Hopp, U.; Schmidt, M.; Ries, C.; Brosch, N. Structure of Brightest Cluster Galaxies and Intracluster Light. {\em \apjs} {\bf 2020}, {\em 247}, 43. [\href{http://dx.doi.org/10.3847/1538-4365/ab733b}{CrossRef}]
\bibitem[Contini; Gu (2020)]{contini20} Contini, E.; Gu, Q. On the Mass Distribution of the Intracluster Light in Galaxy Groups and Clusters. {\em \apj} {\bf 2020}, {\em 901}, 128. [\href{http://dx.doi.org/10.3847/1538-4357/abb1aa}{CrossRef}]
\bibitem[Contini; Gu (2021)]{contini21} Contini, E.; Gu, Q. Brightest Cluster Galaxies and Intracluster Light: Their Mass Distribution in the Innermost Regions of Groups and Clusters. {\em \apj} {\bf 2021}, {\em 915}, 106. [\href{http://dx.doi.org/10.3847/1538-4357/ac01e6}{CrossRef}]
\bibitem[Sampaio-Santos et al. (2021)]{sampaio21} Sampaio-Santos, H.; Zhang, Y.; Ogando, R.L.C.; Shin, T.; Golden-Marx, J.B.; Yanny, B.; Herner, K.; Hilton, M.; Choi, A.; Gatti, M.; et al. Is Diffuse Intracluster Light a Good Tracer of the Galaxy Cluster Matter Distribution? {\em \mnras} {\bf 2021}, {\em 501} 1300--1315. [\href{http://dx.doi.org/10.1093/mnras/staa3680}{CrossRef}]
\bibitem[Poliakov et al. (2021)]{poliakov21} Poliakov, D.; Mosenkov A.V.; Brosch, N.; Koriski, S.; Rich, R.M. Quantified Diffuse Light in Compact Groups of Galaxies. {\em \mnras} {\bf 2021}, {\em 503}, 6059--6077. [\href{http://dx.doi.org/10.1093/mnras/stab853}{CrossRef}]
\bibitem[Deason et al. (2021)]{deason21} Deason, A.J.; Oman, K.A.; Fattahi, A.; Schaller, M.; Jauzac, M.; Zhang, Y.; Montes, M.; Bahé, Y.M.; Dalla Vecchia, C.; Kay, S.T.; et al. Stellar Splashback: The Edge of the Intracluster Light. {\em \mnras} {\bf 2021}, {\em 500}, 4181--4192. [\href{http://dx.doi.org/10.1093/mnras/staa3590}{CrossRef}]
\bibitem[Gonzalez et al. (2021)]{gonzalez21} Gonzalez, A.H.; George, T.; Connor, T.; Deason, A.; Donahue, M.; Montes, M.; Zabludoff, A.I.; Zaritsky, D. Discovery of a Possible Splashback Feature in the Intracluster Light of MACS J1149.5+2223. {\em arXiv} {\bf 2021}, {arXiv:2104.04306}.
\bibitem[Lotz et al. (2017)]{lotz17} Lotz, J.M.; Koekemoer, A.; Coe, D.; Grogin, N.; Capak, P.; Mack, J.; Anderson, J.; Avila, R.; Barker, E.A.; Borncamp, D.; et al. The Frontier Field: Survey Design and Initial Results. {\em \apj} {\bf 2017}, {\em 837}, 97. [\href{http://dx.doi.org/10.3847/1538-4357/837/1/97}{CrossRef}]
\bibitem[Hou et al.(2017)]{hou17} Hou, M.; Li, Z.; Peng, E.W.; Liu, C.   Chandra Detection of Intracluster X-Ray sources in Virgo. {\em \apj} {\bf 2017}, {\em 846}, 126. [\href{http://dx.doi.org/10.3847/1538-4357/aa8635}{CrossRef}]
\bibitem[Jin et al.(2019)]{jin19} Jin, X.; Hou, M.; Zhu, Z.; Li, Z.  Chandra Detection of Intracluster X-Ray Sources in Fornax. {\em \apj} {\bf 2019}, {\em 876}, 53. [\href{http://dx.doi.org/10.3847/1538-4357/ab064f}{CrossRef}]

\bibitem[Ragusa et al. (2021)]{ragusa21} Ragusa, R.; Spavone, M.; Iodice, E.; Brough, S.; Raj, M.A.; Paolillo, M.; Cantiello, M.; Forbes, D.A.; La Marca, A.; Ago, G.D.; et al. VEGAS: A VST Early-type Galaxy Survey. VI. The diffuse Light in HCG 86 from the Ultra-Deep VEGAS Images. {\em arXiv} {\bf 2021}, {arXiv:2105.06970}.
\bibitem[Mihos et al. (2017)]{mihos17} Mihos, J.C.; Harding, P.; Feldmeier, J.; Rudick, C.; Janowiecki, S.; Morrison, H.; Slater, C.; Watkins, A.  The Burrell Schmidt Deep Virgo Survey: Tidal Debris, Galaxy Halos, and Diffuse Intracluster Light in the Virgo Cluster. {\em \apj} {\bf 2017}, {\em 834}, 16. [\href{http://dx.doi.org/10.3847/1538-4357/834/1/16}{CrossRef}]
\newpage
\bibitem[Cantiello et al. (2018)]{cantiello18} Cantiello, M.; D'Abrusco, R.; Spavone, M.; Paolillo, M.; Capaccioli, M.; Limatola, L.; Grado, A.; Iodice, E.; Raimondo, G.; Napolitano, N.; et al.   VEGAS-SSS. II. Comparing the Globular Cluster Systems in NGC3115 and NGC1399 Using VEGAS and FDS Survey Data. The Quest for a Common Genetic Heritage of Globular Cluster System. {\em \aap} {\bf 2018}, {\em 611}, A93. [\href{http://dx.doi.org/10.1051/0004-6361/201730649}{CrossRef}]
\bibitem[Gonzalez et al. (2007)]{gonzalez07} Gonzalez, A.H.; Zaritsky, D.; Zabludoff, A.I. A Census of Baryons in Galaxy Groups and Clusters. {\em \apj} {\bf 2007}, {\em 666}, 147--155. [\href{http://dx.doi.org/10.1086/519729}{CrossRef}]
\bibitem[Iodice et al. (2016)]{iodice16} Iodice, E.; Capaccioli, M.; Grado, A.; Limatola, L.; Spavone, M.; Napolitano, N.R.; Paolillo, M.; Peletier, R.F.; Cantiello, M.; Lisker, T.; et al. The Fornax Deep Survey with VST. I. The Extended and Diffuse Stellar Halo of NGC1399 out to 192 kpc. {\em \apj} {\bf 2016}, {\em 820}, 42. [\href{http://dx.doi.org/10.3847/0004-637X/820/1/42}{CrossRef}]
\bibitem[Postman et al. (2012)]{postman12} Postman, M.; Coe, D.; Benitez, N.; Bradley, L.; Broadhurst, T.; Donahue, M.; Ford, H.; Graur, O.; Graves, G.; Jouvel, S.; et al. The Cluster Lensing and Supernova Survey with Hubble: An Overview. {\em  Astrophys. J. Suppl. Ser.} {\bf 2012}, {\em 199}, 25. [\href{http://dx.doi.org/10.1088/0067-0049/199/2/25}{CrossRef}]
\bibitem[Peng et al. (2002)]{peng02} Peng, C.Y.; Ho, L.C.; Impey, C.D.; Rix, H.W.  Detailed Structural Decomposition of Galaxy Images. {\em \apj} {\bf 2002}, {\em 124}, 266--293. [\href{http://dx.doi.org/10.1086/340952}{CrossRef}]
\bibitem[Navarro et al. (1997)]{navarro97} Navarro, J.F.; Frenk, C.S.; White, S.D.M. A Universal Density Profile from Hierarchical Clustering. {\em \apj} {\bf 1997}, {\em 490}, 493--508. [\href{http://dx.doi.org/10.1086/304888}{CrossRef}]
\bibitem[Jaffe W. (1983)]{jaffe83} Jaffe, W. A Simple Model for the Distribution of Light in Spherical Galaxies. {\em \mnras} {\bf 1983}, {\em 202}, 995--999. [\href{http://dx.doi.org/10.1093/mnras/202.4.995}{CrossRef}]
\bibitem[Da Rocha et al. (2008)]{darocha08} Da Rocha, C.; Ziegler, B.L.; Mendes de Oliveira, C. Intragroup Diffuse Light in Compact Groups of Galaxies-II. HCG 15, 35 and 51. {\em \mnras} {\bf 2008}, {\em 388}, 1433--1443. [\href{http://dx.doi.org/10.1111/j.1365-2966.2008.13500.x}{CrossRef}]
\bibitem[Guennou et al. (2012)]{guennou12} Guennou, L.; Adami, C.; Da Rocha, C; Durret, F.; Ulmer, M.P.; Allam, S.; Basa, S.; Benoist, C.; Biviano, A.; Clowe, D.; et al.  Intracluster Light in Clusters of Galaxies at Redshifts $0.4<z<0.8$. {\em \aap} {\bf 2012}, {\em 537}, A64.
\bibitem[Adami et al. (2013)]{adami13} Adami, C.; Durret, F.; Guennou, L.; Rocha, C. D.   Diffuse Light in the Young Cluster of Galaxies CL J1449+0856 at $z=2.07$. {\em \aap} {\bf 2013}, {\em 551}, A20. [\href{http://dx.doi.org/10.1051/0004-6361/201220282}{CrossRef}]
\bibitem[Ellien et al. (2019)]{ellien19} Ellien, A.; Durret, F.; Adami, C.; Martinet, N.; Lobo, C.; Jauzac, M. The Complex Case of MACS J0717.5+3745 and its Extended Filament: Intracluster Light, Galaxy Luminosity Function, and Galaxy orientations. {\em \aap} {\bf 2019}, {\em 628}, A34. [\href{http://dx.doi.org/10.1051/0004-6361/201935673}{CrossRef}]
\bibitem[Ellien et al. (2021)]{ellien21} Ellien, A.; Slezak, E.; Martinet, N.; Durret, F.; Adami, C.; Gavazzi, R.; Rabaça, C.R.; Rocha, C.D.; Pereira, D.E.   DAWIS, a Detection Algorithm with Wavelets for Intracluster Light Studies. {\em \aap} {\bf 2021}, {\em 649}, A38. [\href{http://dx.doi.org/10.1051/0004-6361/202038419}{CrossRef}]
\bibitem[Murante et al. (2004)]{murante04} Murante, G.; Arnaboldi, M.; Gerhard, O.; Borgani, S.; Cheng, L.M.; Diaferio, A.; Dolag, K.; Moscardini, L.; Tormen, G.; Tornatore, L.; et al. The Diffuse Light in Simulations of Galaxy Clusters. {\em \apj} {\bf 2004}, {\em 607}, L83--L86. [\href{http://dx.doi.org/10.1086/421348}{CrossRef}]
\bibitem[Sommer-Larsen et al. (2005)]{sommer-larsen05} Sommer-Larsen, J.; Romeo, A.D.; Portinari, L. Simulating Galaxy Clusters-III. Properties of the Intraclusters Stars. {\em \mnras} {\bf 2005}, {\em 357}, 478--488. [\href{http://dx.doi.org/10.1111/j.1365-2966.2005.08599.x}{CrossRef}]
\bibitem[Villalobos et al. (2012)]{villalobos12} Villalobos, A.; De Lucia, G.; Borgani, S.; Murante, G. Simulating the Evolution of Disc Galaxies in a Group Environment -I. The Influence of the Global Tidal Field. {\em \mnras} {\bf 2012}, {\em 424}, 2401--2428. [\href{http://dx.doi.org/10.1111/j.1365-2966.2012.20667.x}{CrossRef}]
\bibitem[Smith et al. (2016)]{smith16} Smith, R.; Choi, H.; Lee, J.; Rhee, J.; Sanchez-Janssen, R.; Sukyoung, K.Y.  The Preferential Tidal Stripping of Dark Matter Versus Stars in Galaxies. {\em \apj} {\bf 2016}, {\em 833}, 109. [\href{http://dx.doi.org/10.3847/1538-4357/833/1/109}{CrossRef}]
\bibitem[Burke et al. (2012)]{burke12} Burke, C.; Collins, C.A.; Stott, J.P.; Hilton, M. Measurements of the Intracluster Light at $z\sim 1$. {\em \mnras} {\bf 2012}, {\em 425}, 2058--2068. [\href{http://dx.doi.org/10.1111/j.1365-2966.2012.21555.x}{CrossRef}]
\bibitem[Oliva-Altamirano et al. (2014)]{oliva14} Oliva-Altamirano, P.; Brough, S.; Lidman, C.; Couch, W.J.; Hopkins, A.M.; Colless, M.; Taylor, E.; Robotham, A.S.G.; Gunawardhana, M.L.P.; Ponman, T.; et al. Galaxy And Mass Assembly (GAMA): Testing Galaxy Formation Models Through the Most Massive Galaxies in the Universe. {\em \mnras} {\bf 2014}, {\em 440}, 762--775. [\href{http://dx.doi.org/10.1093/mnras/stu277}{CrossRef}]
\bibitem[Zhang et al. (2016)]{zhang16} Zhang, Y.; Miller, C.; McKay, T.; Rooney, P.; Evrard, A.E.; Romer, A.K.; Perfecto, R.; Song, J.; Desai, S.; Mohr, J.; et al. Galaxies in X-Ray Selected Clusters and Groups in Dark Energy Survey Data. I. Stellar Mass Growth of Bright Central Galaxies since $z\sim 1.2$. {\em \apj} {\bf 2016}, {\em 816}, 98. [\href{http://dx.doi.org/10.3847/0004-637X/816/2/98}{CrossRef}]
\bibitem[Han et al. (2018]{han18} Han, S.; Smith, R.; Choi, H.; Cortese, L.; Catinella, B.; Contini, E.; Sukyoung, K.Y.  YZiCS: Preprocessing of Dark Halos in the Hydrodynamic Zoom-in Simulation of Clusters. {\em \apj} {\bf 2018}, {\em 866}, 78.
\bibitem[Contini et al. (2017)]{contini17a} Contini, E.; Kang, X.; Romeo, A.D.; Xia, Q. Constraints on the Evolution of the Galaxy Stellar Mass Function I: Role of Star Formation, Mergers, and Stellar Stripping. {\em \apj} {\bf 2017}, {\em 837}, 27. [\href{http://dx.doi.org/10.3847/1538-4357/aa5d16}{CrossRef}]
\bibitem[Contini et al. (2017)]{contini17b} Contini, E.; Kang, X.; Romeo, A.D.; Xia, Q.; Yi, S.K. Constraints on the Evolution of the Galaxy Stellar Mass Function. II. The Quenching Timescale of Galaxies and its Implication for Their Star Formation Rates. {\em \apj} {\bf 2017}, {\em 849}, 156. [\href{http://dx.doi.org/10.3847/1538-4357/aa93dd}{CrossRef}]
\bibitem[Chandrasekhar S. (1943)]{chandrasekhar43} Chandrasekhar, S. Stochastic Problems in Physics and Astronomy. {\em Rev. Mod. Phys.} {\bf 1943}, {\em 15}, 1--89. [\href{http://dx.doi.org/10.1103/RevModPhys.15.1}{CrossRef}]
\bibitem[Contini et al. (2012)]{contini12} Contini, E.; De Lucia, G.; Borgani, S. Statistic of Substructures in Dark Matter Haloes. {\em \mnras} {\bf 2012}, {\em 420}, 2978--2989. [\href{http://dx.doi.org/10.1111/j.1365-2966.2011.20149.x}{CrossRef}]
\bibitem[Roberts et al. (2015)]{roberts15} Roberts, I.D.; Parker, L.C.; Joshi, G.D.; Evans, F.A. Mass-Segregation Trends in SDSS Galaxy Groups. {\em \mnras} {\bf 2015}, {\em 448}, L1--L5. [\href{http://dx.doi.org/10.1093/mnrasl/slu188}{CrossRef}]
\bibitem[Contini et al. (2015)]{contini15} Contini, E.; Kang, X. Semi-Analytic Predictions of the Mass Segregation from Groups to Clusters. {\em \mnras} {\bf 2015}, {\em 453}, L53--L57. [\href{http://dx.doi.org/10.1093/mnrasl/slv103}{CrossRef}]
\bibitem[Kim et al. (2020)]{kim20} Kim, S.; Contini, E.; Choi, H.; Han, S.; Lee, J.; Oh, S.; Kang, X.; Sukyoung, K.Y.   YZiCS: On the Mass Segregation of Galaxies in Clusters. {\em \apj} {\bf 2020}, {\em 905}, 12. [\href{http://dx.doi.org/10.3847/1538-4357/abbfa6}{CrossRef}]
\bibitem[Gao et al. (2011)]{gao11} Gao, L.; Frenk, C.S.; Boylan-Kolchin, M.; Jenkins, A.; Springel, V.; White, S.D.M.  The Statistics of the Subhalo Abundance of Dark Matter Haloes. {\em \mnras} {\bf 2008}, {\em 410}, 2309--2314. [\href{http://dx.doi.org/10.1111/j.1365-2966.2010.17601.x}{CrossRef}]
\bibitem[Prada et al. (2012)]{prada12} Prada, F.; Klypin, A.A.; Cuesta, A.J.; Betancort-Rijo, J.E. ; Primack, J.   Halo Concentration in the Standard $\Lambda$ Cold Dark Matter Cosmology. {\em \mnras} {\bf 2012}, {\em 423}, 3018--3030.
\bibitem[Binney et al. (2008)]{binney08} Binney, J.; Tremaine, S. {\em Galactic Dynamics: Second Edition}; Princeton University Press: Princeton, NJ,  USA, {2008}.
\bibitem[Ko; Jee (2018)]{ko18} Ko, J.; Jee, M.J. Evidence for the Existence of Abundant Intracluster Light at $z=1.24$. {\em \apj} {\bf 2018}, {\em 862}, 95. [\href{http://dx.doi.org/10.3847/1538-4357/aacbda}{CrossRef}]
\bibitem[Spavone et al. (2018)]{spavone18} Spavone, M.; Iodice, E.; Capaccioli, M.; Bettoni, D.; Rampazzo, R.; Brosch, N.; Cantiello, M.; Napolitano, N.R.; Limatola, L.; Grado, A.; et al. VEGAS: A VST Early-type Galaxy Survey. III. Mapping the Galaxy Structure, Interactions, and Intragroup Light in the NGC5018 Group. {\em \apj} {\bf 2018}, {\em 864}, 149. [\href{http://dx.doi.org/10.3847/1538-4357/aad6e9}{CrossRef}]
\bibitem[Henden et al. (2019)]{henden19} Henden, N.A.; Puchwein, E.; Sijacki, D. The Redshift Evolution of X-ray and Sunyaev-Zel'dovich Scaling Relations in the FABLE Simulations. {\em \mnras} {\bf 2019}, {\em 489}, 2439--2470. [\href{http://dx.doi.org/10.1093/mnras/stz2301}{CrossRef}]
\bibitem[Canas et al. (2020)]{canas20} Canas, R.; Lagos, C.; Elahi, P.J.; Power, C.; Welker, C.; Dubois, Y.; Pichon, C.   From Stellar Haloes to Intracluster Light: The Physics of the Intra-Halo Stellar Component in Cosmological Hydrodynamical Simulations. {\em \mnras} {\bf 2020}, {\em 494}, 4314--4333. [\href{http://dx.doi.org/10.1093/mnras/staa1027}{CrossRef}]
\bibitem[Lin; Mohr (2004)]{lin04} Lin, Y.T.; Mohr, J. K-band Properties of Galaxy Clusters and Groups: Brightest Cluster Galaxies and Intracluster Light. {\em \apj} {\bf 2004}, {\em 617}, 879--895. [\href{http://dx.doi.org/10.1086/425412}{CrossRef}]
\bibitem[De Lucia; Blaizot (2007)]{delucia07} De Lucia, G.; Blaizot, J. The Hierarchical Formation of the Brightest Cluster Galaxies. {\em \mnras} {\bf 2007}, {\em 375}, 2--14. [\href{http://dx.doi.org/10.1111/j.1365-2966.2006.11287.x}{CrossRef}]
\bibitem[Lee; Yi (2017)]{lee17} Lee, J.; Yi, S.K. Formation and Assembly History of Stellar Components in Galaxies as a Function of Stellar and Halo Mass. {\em \apj} {\bf 2017}, {\em 836}, 161. [\href{http://dx.doi.org/10.3847/1538-4357/aa5b87}{CrossRef}]
\bibitem[Lin et al. (2013)]{lin13} Lin, Y.T.; Brodwin, M.; Gonzalez, A.H.; Bode, P.; Eisenhardt, P.R.; Stanford, S.A.; Vikhlinin, A. The Stellar Mass Growth of Brightest Cluster Galaxies in the IRAC Shallow Cluster Survey. {\em \apj} {\bf 2013}, {\em 771}, 61. [\href{http://dx.doi.org/10.1088/0004-637X/771/1/61}{CrossRef}]
\bibitem[Lidman et al. (2012)]{lidman12} Lidman, C.; Suherli, J.; Muzzin, A.; Wilson, G.; Demarco, R.; Brough, S.; Rettura, A.; Cox, J.; DeGroot, A.; Yee, H.K.C.; et al.   Evidence for Significant Growth in the Stellar Mass of Brightest Cluster Galaxies Over the Past 10 Billion Years. {\em \mnras} {\bf 2012}, {\em 427}, 550--568. [\href{http://dx.doi.org/10.1111/j.1365-2966.2012.21984.x}{CrossRef}]
\bibitem[Jee (2010)]{jee10} Jee, M.J. Tracing the Peculiar Dark Matter Structure in the Galaxy Cluster CI0024+17 with Intracluster Stars and Gas. {\em \apj} {\bf 2010}, {\em 717}, 420--434. [\href{http://dx.doi.org/10.1088/0004-637X/717/1/420}{CrossRef}]
\bibitem[Diemer; Kravtsov (2014)]{diemer14} Diemer, B.; Kravtsov, A.V. Dependence of the Outer Profiles of Halos on Their Mass Accretion Rate. {\em \apj} {\bf 2014}, {\em 789}, 1. [\href{http://dx.doi.org/10.1088/0004-637X/789/1/1}{CrossRef}]












\end{thebibliography}


\end{document}